\documentclass[manuscript,screen,nonacm]{acmart}

\usepackage{booktabs}
\usepackage{tabularx}
\usepackage{graphicx}
\usepackage{balance}
\usepackage{xcolor}
\definecolor{cbblue}{RGB}{0,114,178}
\definecolor{cborange}{RGB}{213,94,0}
\definecolor{cbyellow}{RGB}{230,159,0}
\definecolor{cblightblue}{RGB}{86,180,233}
\usepackage{tikz}
\usepackage{pgfplots}
\usepackage{float}
\usepackage{subcaption}
\pgfplotsset{compat=1.18}
\usetikzlibrary{shapes.geometric, arrows.meta, positioning, calc, patterns, decorations.pathreplacing}

\AtBeginDocument{%
  }

\setcopyright{acmcopyright}
\copyrightyear{2026}
\acmYear{2026}
\acmDOI{XXXXXXX.XXXXXXX}

\acmConference[FAccT '26]{ACM Conference on Fairness, Accountability, and Transparency}{June 25--28, 2026}{Montr\'{e}al, Canada}
\acmISBN{978-1-4503-XXXX-X/26/06}

\begin{document}

\title{Beyond Explanation: Evidentiary Rights for Algorithmic Accountability}

\author{Matthew Stewart}
\affiliation{%
  \institution{Independent Researcher}
  \country{USA}
}

\begin{abstract}
Algorithmic accountability scholarship has focused heavily on explanation, helping affected parties understand why decisions were made. We argue this focus is insufficient. Explanation without evidentiary access does not enable meaningful contestation. A person told ``your risk score was 0.73'' understands the decision but cannot verify the score, test alternatives, or produce counter-evidence. We introduce a taxonomy of contestation failures, showing that most accountability interventions address only one failure mode (opacity) while leaving four others unaddressed. Drawing on analysis of 168 legal cases spanning algorithmic decision-making contexts, we find that contestation faces a \emph{two-gate} structure: a procedural gate (evidentiary access) and a doctrinal gate (substantive liability rules). Among litigated cases, those without evidence access almost never succeed (9\%); those with access succeed at rates approaching 97\% in domains without liability shields. Where doctrinal immunities apply (e.g., Section 230), even full evidentiary scrutiny produces no liability. This association almost certainly reflects selection effects; our empirical contribution is diagnostic rather than causal. The data identify where contestation fails among observable cases, not whether providing access would change outcomes for currently-excluded cases. We propose \emph{evidentiary rights} as the missing procedural component, and develop counterfactual interrogation rights that allow affected parties to probe decision systems with modified inputs and observe whether outcomes change, without requiring disclosure of model internals. This reframes algorithmic accountability from a transparency problem to a procedural rights problem.
\end{abstract}



\maketitle

\section{Introduction}
\label{sec:introduction}

\noindent Consider a loan applicant denied credit by an algorithmic system. She receives an explanation: ``Your application was denied due to insufficient credit history and high debt-to-income ratio.'' This satisfies explanation requirements, but can she \emph{contest} the decision? Contestation requires more than understanding. She would need to verify her data was accurately assessed, test whether changes would alter the outcome, and compare her treatment to similar applicants. Current frameworks provide none of this, so she understands the decision but cannot challenge it. This paper argues that algorithmic accountability scholarship has conflated \emph{explanation} (why a decision was made) with \emph{contestation} (the ability to challenge whether it was correct). You can have perfect explanation with zero contestation capability, which is precisely the situation most frameworks create.
\begin{table}[!ht]
\caption{Explanation vs. Contestation: Distinct Functions}
\label{tab:explanation-contestation}
\small
\centering
\renewcommand{\arraystretch}{1.2}
\begin{tabular}{@{}l@{\hspace{2em}}l@{\hspace{2em}}l@{}}
\toprule
\textbf{Dimension} & \textbf{Explanation} & \textbf{Contestation} \\
\midrule
Core question & ``Why this decision?'' & ``Was it correct?'' \\
Information flow & One-way (org → person) & Interactive (person probes) \\
Control & Organization selects content & Affected party directs inquiry \\
Evidence type & Rationale provided & Testable claims verified \\
Assumes & Decision is final & Decision may be wrong \\
Success metric & Recipient understands & Recipient can challenge \\
Current coverage & High (XAI, GDPR, FCRA) & Low (our contribution) \\
\bottomrule
\end{tabular}
\end{table}

\textbf{Situating the proposal.} Legal systems already provide mechanisms for challenging decisions with evidence. Litigation discovery allows parties to compel document production, and administrative procedures require agencies to disclose reasoning. But these mechanisms activate \emph{after} formal proceedings begin. Our proposal addresses the absence of evidentiary access \emph{before} litigation, in the administrative appeals, workplace disputes, and platform contexts where most algorithmic contestation occurs. Counterfactual interrogation is not a reinvention of discovery but a pre-litigation mechanism that enables affected parties to generate evidence without filing a lawsuit or hiring counsel. Evidentiary rights can attach to existing enforcement frameworks such as Fair Housing Act testing, CFPB supervision, and GDPR investigations. Where no such framework exists, new legislation or regulatory reinterpretation is needed.

\textbf{The gap.} Prior work recognizes limits of explanation-focused approaches~\cite{edwards2017slave,mulligan2019shaping,mulligan2020contestability,kaminski2021right}. We build on this foundation with three contributions: (1) a taxonomy of \emph{failure modes} showing where contestation breaks down; (2) \emph{empirical evidence} from 168 legal cases revealing a two-gate structure where evidence access is a precondition but not a guarantee; and (3) \emph{operational specification} of counterfactual interrogation as a procedural mechanism.

\textbf{The evidence.} Our survey of 168 legal cases across employment, housing, healthcare, criminal justice, government benefits, credit/insurance, and platform liability in 19 jurisdictions reveals a two-gate structure. At the \emph{procedural gate}, without evidence access, contestation almost never succeeds, with only 1 of 11 no-access cases (9\%) achieving favorable outcomes. At the \emph{doctrinal gate}, outcomes depend on substantive liability rules. In domains without immunity shields, success rates approach 97\% (56 of 58 resolved partial-access cases). But where doctrinal immunities apply, most notably Section 230 for platforms, even full evidentiary examination produces no liability. All six platform cases had full algorithmic scrutiny, and all six were denied on doctrinal grounds. This demonstrates that evidentiary access \emph{enables but does not ensure} accountability. Our contribution is diagnostic. We identify where contestation systematically fails, not that access causes success. The normative argument rests on procedural fairness, that affected parties deserve the opportunity to present evidence, rather than on causal claims about outcomes. To address this, we propose \emph{counterfactual interrogation rights}, which allow affected parties to test decisions with modified inputs, observe whether outcomes change, and generate evidence for appeals, all without disclosing model internals.

\textbf{Scope and normative grounding.} We focus on consequential algorithmic decisions that produce material adverse effects, including denial of employment, credit, or housing, as well as content removal and account termination. Our normative commitment is that affected parties should be able to meaningfully challenge such decisions~\cite{tyler2003procedural,binns2018algorithmic}. This requires defense, since these domains involve private actors. Three grounds support imposing such obligations. First, \emph{existing regulatory precedent}: FCRA already requires adverse action disclosures and dispute mechanisms, and ECOA and ADA mandate similar protections. Evidentiary rights extend existing regimes, not create novel obligations. Second, \emph{reliance and power asymmetry}: when organizations deploy algorithmic systems affecting employment, credit, or housing, affected parties have limited alternatives and substantial reliance interests justifying procedural protections. Third, \emph{error correction benefits organizations}: evidentiary access reveals system defects otherwise undiscovered, improving decision quality and reducing liability exposure. These grounds do not establish unlimited rights, and graduated implementation (Section~\ref{sec:regulatory-comparison}) calibrates access to decision stakes.

\textbf{Why algorithms specifically.} Algorithmic decisions warrant evidentiary rights for reasons beyond those justifying scrutiny of human decisions. The first is \emph{scale and replicability}. Algorithms make identical decisions across thousands of cases, creating verifiable patterns that individual human decisions do not. The second is \emph{interrogation limits}. Human decisionmakers can be deposed and cross-examined, but algorithms produce no testimony, and without testing access, affected parties have no way to probe the reasoning behind a decision. The third is \emph{precision}. Algorithmic scores such as 0.73 against a threshold of 0.80 permit exact counterfactual testing in ways that human judgments rarely allow. These features make algorithmic decisions both more amenable to evidentiary testing and more in need of it. Importantly, this does not require disclosing proprietary internals. Counterfactual interrogation provides behavioral testing, not model disclosure, and graduated implementation calibrates access so that high-risk domains warrant mediated access via licensed representatives.

\subsection{Related Work}

We accept that explanation serves legitimate functions~\cite{lipton2018mythos,miller2019explanation} but argue it is insufficient for contestation, providing empirical evidence for concerns raised by Selbst and Barocas~\cite{selbst2018intuitive}. Critical algorithm studies document how algorithmic systems reproduce structural inequalities~\cite{eubanks2018automating,noble2018algorithms,benjamin2019race}. Hoffmann~\cite{hoffmann2019fairness} argues that individual-focused remedies cannot address structural harms, and we agree that evidentiary rights complement rather than replace structural interventions.

Our work builds most directly on two lines of scholarship. Mulligan et al.~\cite{mulligan2019shaping,mulligan2020contestability} argue that contestability requires more than transparency, an insight developed further in HCI research~\cite{lyons2021conceptualising,alfrink2023contestable}. Kaminski and Malgieri~\cite{kaminski2021right} analyze the GDPR ``right to contest'', and we complement their doctrinal analysis with empirical evidence and an implementation mechanism. Algorithmic auditing~\cite{raji2020closing,vecchione2021algorithmic,raji2022outsider} operates at the population level, while our proposal enables individual contestation. Counterfactual fairness~\cite{kusner2017counterfactual} defines fairness criteria, and we specify a procedural mechanism for enforcing them. Finally, Selbst et al.~\cite{selbst2019fairness}, Edwards and Veale~\cite{edwards2017slave}, Power~\cite{power1997audit}, and Edelman~\cite{edelman2016working} caution that formal mechanisms can become ritualistic. We engage these critiques directly in Section~\ref{sec:limitations}.
\section{A Taxonomy of Contestation Failures}
\label{sec:taxonomy}

We define \emph{contestation failure} as any structural barrier that prevents an affected party from mounting a meaningful challenge to an algorithmic decision. This includes not only the absence of process but also processes that are not rationally designed to permit challenge. A forum with unrealistic standards or no remedies constitutes failure even if formally available. The following taxonomy is our original analytical contribution, organizing five distinct failure modes, each requiring different interventions.

\smallskip
\noindent\textbf{Type 1: Opacity.} The affected party does not know why the decision was made. For example, an application is rejected with no reason given. Existing explanation requirements such as GDPR Article 22 and adverse action notices address this failure mode, and it is the primary focus of current accountability research.

\smallskip
\noindent\textbf{Type 2: Explanation Without Evidence.} The affected party receives an explanation but cannot verify or challenge it. For example, an applicant told ``insufficient credit history'' cannot verify the assessment's accuracy or compare her treatment to similar applicants. The boundary between Types 1 and 2 is porous, as a legally sufficient explanation may satisfy regulatory requirements without producing genuine understanding~\cite{miller2019explanation}. This strengthens our argument, because even assuming perfect understanding, contestation still requires evidentiary access. Throughout this paper, when we refer to an affected party ``understanding'' a decision, we mean they received a legally compliant explanation, not that they fully comprehend the system's operation. This is the failure mode we address.

\smallskip
\noindent\textbf{Type 3: Evidence Without Forum.} The affected party can access evidence but has no venue for challenge. For example, content is removed with a timestamp provided but no appeal mechanism exists. The intervention is mandatory appeal pathways.

\smallskip
\noindent\textbf{Type 4: Forum Without Standards.} Appeals exist but with no clear success criteria, and reviewers have unbounded discretion. The intervention is published standards and precedent systems.

\smallskip
\noindent\textbf{Type 5: Standards Without Remedies.} A successful challenge produces no correction. For example, a bureau acknowledges an error but does not update the record. The intervention is mandatory remediation.

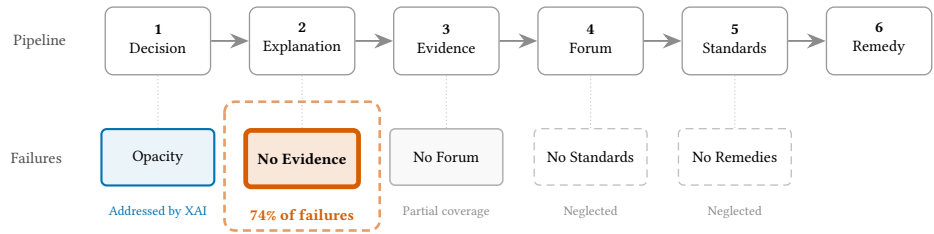
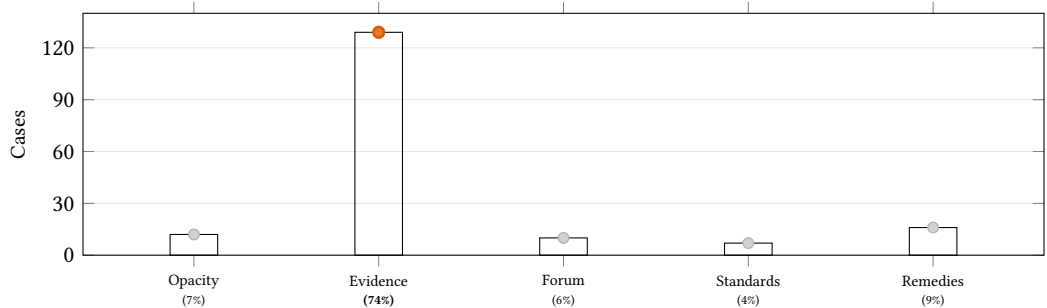
\begin{figure}[ht]
\centering
\begin{subfigure}[b]{\columnwidth}
\centering
\begin{tikzpicture}[
    node distance=0.6cm and 0.5cm,
    stage/.style={rectangle, rounded corners=3pt, draw=black!40, fill=white, minimum width=1.4cm, minimum height=0.9cm, align=center, font=\scriptsize, line width=0.5pt},
    failure/.style={rectangle, rounded corners=2pt, minimum width=1.5cm, minimum height=0.75cm, align=center, font=\scriptsize},
    addressed/.style={failure, draw=cbblue, fill=cbblue!8, line width=0.8pt},
    critical/.style={failure, draw=cborange, fill=cborange!15, line width=1.8pt, font=\scriptsize\bfseries},
    partial/.style={failure, draw=black!30, fill=gray!5, line width=0.6pt},
    neglected/.style={failure, draw=black!25, fill=white, line width=0.5pt, densely dashed},
    arrow/.style={-{Stealth[length=2.5mm, width=2mm]}, line width=0.7pt, black!45},
    connector/.style={line width=0.4pt, black!25, densely dotted},
]

\node[stage] (s1) {\textbf{1}\\\smallskip Decision};
\node[stage, right=of s1] (s2) {\textbf{2}\\\smallskip Explanation};
\node[stage, right=of s2] (s3) {\textbf{3}\\\smallskip Evidence};
\node[stage, right=of s3] (s4) {\textbf{4}\\\smallskip Forum};
\node[stage, right=of s4] (s5) {\textbf{5}\\\smallskip Standards};
\node[stage, right=of s5] (s6) {\textbf{6}\\\smallskip Remedy};

\draw[arrow] (s1) -- (s2);
\draw[arrow] (s2) -- (s3);
\draw[arrow] (s3) -- (s4);
\draw[arrow] (s4) -- (s5);
\draw[arrow] (s5) -- (s6);

\node[left=0.4cm of s1, font=\scriptsize, text=black!70, align=right] {Pipeline};

\node[addressed, below=0.7cm of s1] (t1) {Opacity};
\node[critical, below=0.7cm of s2] (t2) {No Evidence};
\node[partial, below=0.7cm of s3] (t3) {No Forum};
\node[neglected, below=0.7cm of s4] (t4) {No Standards};
\node[neglected, below=0.7cm of s5] (t5) {No Remedies};

\node[left=0.4cm of t1, font=\scriptsize, text=black!70, align=right] {Failures};

\draw[connector] (s1.south) -- (t1.north);
\draw[connector] (s2.south) -- (t2.north);
\draw[connector] (s3.south) -- (t3.north);
\draw[connector] (s4.south) -- (t4.north);
\draw[connector] (s5.south) -- (t5.north);

\node[below=0.15cm of t1, font=\tiny, text=cbblue] {Addressed by XAI};
\node[below=0.15cm of t2, font=\scriptsize\bfseries, text=cborange] {74\% of failures};
\node[below=0.15cm of t3, font=\tiny, text=black!45] {Partial coverage};
\node[below=0.15cm of t4, font=\tiny, text=black!45] {Neglected};
\node[below=0.15cm of t5, font=\tiny, text=black!45] {Neglected};

\draw[cborange!60, line width=1pt, rounded corners=4pt, densely dashed]
    ([xshift=-0.25cm, yshift=0.35cm]t2.north west) rectangle ([xshift=0.25cm, yshift=-0.55cm]t2.south east);

\end{tikzpicture}
\caption{Contestation pipeline with failure points. Each stage requires the previous; failure at any point blocks contestation.}
\label{fig:taxonomy-pipeline}
\end{subfigure}

\vspace{0.2cm}

\begin{subfigure}[b]{\columnwidth}
\centering
\begin{tikzpicture}
\begin{axis}[
    ybar,
    bar width=18pt,
    width=0.95\columnwidth,
    height=4.8cm,
    ylabel={Cases},
    ylabel style={font=\small},
    symbolic x coords={Opacity, Evidence, Forum, Standards, Remedies},
    xtick=data,
    xticklabels={Opacity\\{\tiny(7\%)}, Evidence\\{\tiny\textbf{(74\%)}}, Forum\\{\tiny(6\%)}, Standards\\{\tiny(4\%)}, Remedies\\{\tiny(9\%)}},
    x tick label style={font=\scriptsize, align=center},
    ymin=0,
    ymax=140,
    ytick={0,30,60,90,120},
    ymajorgrids=true,
    grid style={line width=0.3pt, gray!20},
    enlarge x limits=0.15,
    nodes near coords,
    every node near coord/.append style={font=\scriptsize},
    scatter/classes={
        gray={fill=gray!35, draw=gray!60, line width=0.5pt},
        orange={fill=cborange!80, draw=cborange, line width=1pt}
    },
]

\addplot[
    scatter,
    scatter src=explicit symbolic,
] table[x=cat, y=val, meta=class] {
cat val class
Opacity 12 gray
Evidence 129 orange
Forum 10 gray
Standards 7 gray
Remedies 16 gray
};

\end{axis}
\end{tikzpicture}
\caption{Distribution among litigated cases by primary failure type (n=168). Evidence access dominates observed cases (74\%); opacity accounts for only 7\%. This reflects observability, not necessarily prevalence.}
\label{fig:taxonomy-empirical}
\end{subfigure}

\caption{Taxonomy of contestation failures: conceptual framework and observed distribution. (a) The contestation pipeline shows where barriers occur; current frameworks address opacity (Type 1) but leave evidentiary access (Type 2) undertheorized. (b) Among cases reaching litigation, evidence access is the most frequently observed barrier: 126 of 168 cases (74\%) involve evidentiary inaccessibility as the primary obstacle. This distribution reflects litigation selection patterns, not necessarily the prevalence of barriers overall---Type 1 failures may not generate litigation, while Types 3--5 failures may be harder to identify in court records.}
\label{fig:taxonomy}
\end{figure}

Current algorithmic accountability research and regulation concentrates heavily on Type 1 (opacity), with interpretable ML, XAI, and explanation requirements all addressing the question ``why was this decision made?'' Types 2--5 receive far less attention (Figure~\ref{fig:taxonomy}). Our legal case analysis (Section~\ref{sec:legal-evidence}) finds that evidence access strongly predicts outcomes among litigated cases. However, the taxonomy is an \emph{analytical framework}, not an empirical claim about where contestation fails most often, and which modes dominate empirically depends on what we observe.

This taxonomy reframes the accountability problem. The question is not ``how do we explain algorithmic decisions?'' but ``how do we enable meaningful contestation of algorithmic decisions?'' Explanation is one component, and evidentiary access, adjudication forums, clear standards, and effective remedies are equally necessary. We focus on Type 2 (evidentiary access) because it is undertheorized and strongly predictive of outcomes among litigated cases. We do not claim Type 2 is the most important barrier overall, as that would require observing failures that never reach litigation. Our empirical contribution is documenting that among observable cases, evidentiary access is strongly associated with success. Types 3--5 are essential complements that we briefly specify in Section~\ref{sec:evidentiary-rights}.


\section{Evidence Access and Accountability Outcomes}
\label{sec:legal-evidence}

We tested whether evidence access predicts accountability outcomes through systematic analysis of 168 legal cases involving algorithmic decision-making across seven domains (employment, credit/insurance, housing, healthcare, criminal justice, government benefits, and platform liability) from 2012 to 2025. Cases span 19 countries and 30 jurisdiction-level units, including US federal and state courts, the EU/CJEU, and national courts and data protection authorities across 17 additional countries. The full case database appears in Appendix~\ref{appendix:cases}.

\textbf{Case identification.} We used a multi-source identification strategy to maximize coverage while acknowledging that no sample of algorithmic accountability cases can be comprehensive.
\begin{itemize}
    \item \emph{Legal databases} (Westlaw, LexisNexis). We searched using terms including ``algorithm,'' ``automated decision,'' ``machine learning,'' ``artificial intelligence,'' ``scoring system,'' ``risk assessment,'' and ``predictive,'' combined with domain terms (``employment,'' ``credit,'' ``housing,'' etc.) and outcome terms (``discrimination,'' ``bias,'' ``due process,'' ``FCRA,'' ``ECOA''). Date range was 2012--2025. We reviewed approximately 800 initial results, excluding cases where algorithms played no material role.
    \item \emph{Regulatory actions}. We reviewed published enforcement actions from EEOC (pattern-or-practice cases involving selection procedures), CFPB (fair lending and FCRA enforcement), FTC (algorithmic deception cases), HUD (fair housing), and DOJ Civil Rights Division. We included consent decrees, settlements exceeding \$50,000, and formal findings.
    \item \emph{Secondary sources}. Academic surveys~\cite{citron2014scored,kroll2017accountable}, investigative journalism (The Markup's ``Machine Bias'' series, ProPublica, Tampa Bay Times), and advocacy databases (ACLU, EFF, AI Now Institute case trackers) identified cases not appearing in legal database searches, particularly pre-litigation settlements and international cases.
\end{itemize}

We included cases where (1) an algorithmic system played a material role in an adverse decision affecting identified individuals or groups, (2) affected parties or their representatives attempted contestation through legal, regulatory, or formal administrative channels, and (3) the case produced a documented outcome or remained ongoing with sufficient procedural history to code. We excluded pure policy advocacy without individual claims, academic studies without legal proceedings, and cases where algorithms were incidental to the dispute.

\textbf{Coding scheme.} All coding was conducted by a single researcher. To assess consistency, 50 cases (29\%) were re-coded after a two-week interval, with temporal stability of 94\% for evidence access (47/50), 96\% for outcome (48/50), and 88\% for explanation quality (44/50). Formal inter-rater reliability with independent coders was not assessed, though coding criteria were explicit and primary sources were consulted for disputed cases. We coded four primary variables with operationalized decision rules for borderline cases.
\begin{itemize}
    \item \emph{Evidence access} (Full/Partial/None). Full access required that plaintiffs obtained input data, system outputs, and either algorithmic logic or comparable cases sufficient to construct counterfactual arguments. Partial access meant plaintiffs obtained outputs or could infer patterns from aggregate data but lacked access to inputs or logic. None meant the system remained opaque throughout proceedings. For borderline cases, if plaintiffs could articulate specific, testable claims about system behavior (e.g., ``the system penalizes applicants over 50''), we coded Partial. If claims remained general (e.g., ``the system is biased''), we coded None.
    \item \emph{Outcome} (Achieved/Denied/Ongoing). Achieved required settlement of \$50,000 or more, judgment for plaintiff, injunctive relief, or formal regulatory finding against defendant. The \$50,000 threshold excludes nuisance settlements while capturing cases where defendants conceded material liability. Denied meant dismissal, summary judgment for defendant, or judgment after trial. Ongoing meant the case remained in active litigation or administrative proceedings.
    \item \emph{Explanation quality} (High/Partial/None). High required that organizations disclosed specific factors with weights, thresholds, or quantified contributions. Partial meant organizations identified factor categories without quantification. None meant no explanation was provided or explanations were purely procedural (``a computer made the decision''). We also coded \emph{actionability}, meaning whether explanations enabled affected parties to generate counter-evidence. High-specificity explanations without actionability (e.g., ``your score was 42'' without threshold or factor disclosure) were coded separately but performed similarly to no explanation.
    \item \emph{Legal resources} (High/Medium/Low). High meant government agency or well-resourced public interest organization. Medium meant private counsel with algorithmic accountability experience or union representation. Low meant solo practitioners, legal aid, or pro~se. \emph{Media attention} was coded binary based on national outlet coverage before resolution.
\end{itemize}

\textbf{Limitations.} Potential missing case types include confidential arbitrations, international cases using non-English terminology, informal regulatory resolutions, and administrative appeals. Despite operationalized rules, some cases required judgment calls, and we report reliability statistics to quantify this uncertainty. Selection effects are substantial and addressed in Section~\ref{sec:interpreting}.

\textbf{Scope notes.} We include platform liability cases (n=6) deliberately as a theoretically motivated test of whether evidentiary access alone suffices for accountability. Their consistent failure despite full access isolates the doctrinal variable, as Section 230 forecloses liability regardless of transparency. We acknowledge n=6 is too small for robust inference and treat these cases as illustrative of the doctrinal gate rather than statistically conclusive. Of 168 cases, 48 (29\%) remain ongoing. All success-rate comparisons are computed on resolved cases only (n=120). The concentration of ongoing cases among no-access (59\%) versus access cases (23\%) suggests evidentiary disputes may cause delay.

\subsection{Results}

\begin{table}[!ht]
\caption{Accountability outcomes by evidence access level (n=168)}
\label{tab:legal-outcomes}
\small
\begin{tabular}{@{}lccccc@{}}
\toprule
\textbf{Evidence Access} & \textbf{Achieved} & \textbf{Denied} & \textbf{Ongoing} & \textbf{Total} & \textbf{Success Rate}$^\dagger$ \\
\midrule
Full & 45 & 6 & 1 & 52 & 88\% (45/51) \\
Partial & 56 & 2 & 30 & 88 & 97\% (56/58) \\
\textbf{Any Access} & \textbf{101} & \textbf{8} & \textbf{31} & \textbf{140} & \textbf{93\% (101/109)} \\
None & 1 & 10 & 17 & 28 & 9\% (1/11) \\
\midrule
\textbf{Total} & 102 & 18 & 48 & 168 & 85\% (102/120) \\
\bottomrule
\end{tabular}
\\[2pt]
{\footnotesize $^\dagger$Success rate = Achieved/(Achieved+Denied), computed on resolved cases only. Total column includes all cases; Success Rate column excludes 48 ongoing cases.}
\end{table}

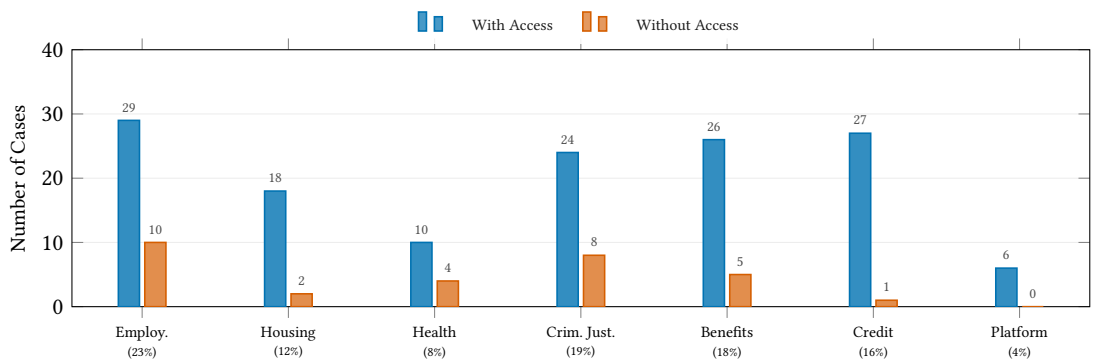
\begin{figure}[h]
\centering
\begin{tikzpicture}
\begin{axis}[
    ybar,
    bar width=8pt,
    width=\columnwidth,
    height=5.0cm,
    ylabel={Number of Cases},
    ylabel style={font=\small},
    symbolic x coords={Employ., Housing, Health, Crim. Just., Benefits, Credit, Platform},
    xtick=data,
    xticklabels={Employ.\\{\tiny(23\%)}, Housing\\{\tiny(12\%)}, Health\\{\tiny(8\%)}, Crim. Just.\\{\tiny(19\%)}, Benefits\\{\tiny(18\%)}, Credit\\{\tiny(16\%)}, Platform\\{\tiny(4\%)}},
    x tick label style={font=\scriptsize, align=center},
    legend style={
        at={(0.5,1.02)},
        anchor=south,
        font=\scriptsize,
        draw=none,
        fill=white,
        fill opacity=0.9,
        legend columns=2,
        column sep=1em
    },
    ymin=0,
    ymax=40,
    ymajorgrids=true,
    grid style={line width=0.3pt, gray!15},
    enlarge x limits=0.08,
    nodes near coords,
    nodes near coords style={font=\tiny, above, yshift=0pt},
    every node near coord/.append style={black!70},
]

\addplot+[
    fill=cbblue!80,
    draw=cbblue,
    line width=0.6pt
] coordinates {(Employ.,29) (Housing,18) (Health,10) (Crim. Just.,24) (Benefits,26) (Credit,27) (Platform,6)};

\addplot+[
    fill=cborange!70,
    draw=cborange,
    line width=0.6pt
] coordinates {(Employ.,10) (Housing,2) (Health,4) (Crim. Just.,8) (Benefits,5) (Credit,1) (Platform,0)};

\legend{With Access, Without Access}

\end{axis}
\end{tikzpicture}
\caption{Case distribution by domain and evidence access level (n=168). Blue bars: cases with evidence access (Full or Partial); orange bars: cases without access. Platform liability cases had full access yet failed due to Section 230 doctrinal immunity.}
\label{fig:domain-outcomes}
\end{figure}

As noted above, our analysis is diagnostic rather than causal. Table~\ref{tab:legal-outcomes} shows the distribution of outcomes by evidence access level. Among resolved cases, those with evidence access succeeded 93\% of the time, compared to just 9\% for those without. This differential almost certainly reflects selection effects (Section~\ref{sec:interpreting}), but it is diagnostically significant. Contestation without evidence access fails at the procedural stage before ever reaching substantive evaluation. Figure~\ref{fig:domain-outcomes} shows this pattern is consistent across all seven domains.

\textbf{The two-gate structure.} An interesting anomaly in the data reveals a deeper pattern. Full-access cases succeed at 88\%, which is lower than partial-access cases at 97\%. The explanation is platform liability. All six full-access failures are cases where Section 230 immunity foreclosed liability despite complete algorithmic scrutiny, and excluding these cases, full-access success returns to 97\%. This reveals what we call accountability's two-gate structure. At the \emph{procedural gate}, the question is whether the affected party can obtain evidence. At the \emph{doctrinal gate}, the question is whether substantive law recognizes the claim. These gates operate sequentially. Without evidence access, cases fail at gate one regardless of their merit. Platform liability cases pass gate one but fail at gate two because Section 230 provides immunity. The critical observation is that cases without access never reach gate two at all. Evidentiary rights would move contestation from an illegitimate filter based on whether you can obtain access to a legitimate one based on whether the evidence supports your claim.

Two cases illustrate this structure. In \emph{EEOC v. iTutorGroup (2023)}~\cite{aboraya2023eeoc}, the first EEOC AI settlement (\$365K), the plaintiffs succeeded because age discrimination was discoverable through aggregated output patterns. The EEOC was able to show that the company's hiring algorithm systematically filtered out applicants over 55, evidence that was only available because output data could be examined across applicants. By contrast, in \emph{State v. Loomis (2016)}~\cite{loomis2016}, the Wisconsin Supreme Court acknowledged that COMPAS opacity was ``a significant problem'' yet upheld its use because Loomis could not verify his score or compare his treatment to similar defendants. Without access to how the risk assessment weighted its inputs, Loomis had no way to challenge the score that influenced his sentence. When evidence access is denied, affected parties are forced into expensive system-level challenges rather than normal individual contestation.

\subsection{Interpreting the Pattern}
\label{sec:interpreting}

The high success rate with access almost certainly reflects selection effects rather than a causal relationship. Five mechanisms may contribute.
\begin{itemize}
    \item \emph{Pre-litigation filtering}. Cases without evidence access may never be filed. Attorneys decline cases they cannot prove, and affected parties without obvious evidence may not recognize viable claims. This likely \emph{understates} the evidence-access effect by excluding cases that failed before our sample.
    \item \emph{Survivorship bias}. Cases that gain access but reveal no wrongdoing may be dropped before resolution, inflating success rates among resolved access cases.
    \item \emph{Settlement dynamics}. Defendants facing damaging discovery often settle regardless of merit, while defendants confident in their position may proceed to judgment. This could inflate or deflate success rates depending on settlement terms.
    \item \emph{Attorney and judicial selection}. Experienced attorneys select cases with higher success probability, and judges may grant discovery to cases they view favorably. Both create correlation between access and merit.
    \item \emph{Alternative causal story}. Case merit may drive both access and outcomes. Strong cases obtain discovery and succeed, with access serving as proxy rather than cause.
\end{itemize}

\textbf{Why the diagnostic finding matters regardless.} We do not claim to resolve these selection issues, as observational litigation data cannot support causal claims. Our argument is procedural, not causal. Currently, cases are filtered at ``can you obtain access?'' before reaching ``does your evidence support your claim?'' The first filter is procedurally illegitimate because it excludes based on litigation resources and organizational opacity rather than case merit. The second filter is legitimate, as cases should fail when evidence does not support the claim. Evidentiary rights would shift filtering to the legitimate basis. The normative argument requires that access is a \emph{procedural precondition} for fair adjudication, not that access causes success.

\textbf{When evidence access was not enough.} To avoid overstating our claims, we also examine cases where evidence access did not lead to success. Among resolved access cases, three resulted in unfavorable outcomes (excluding platform liability). \emph{Redfin steering} involved ambiguous output data, \emph{Murphy v. Macy's} had an insufficient sample size, and \emph{LinkedIn salary estimation} produced legitimate non-discriminatory explanations. These confirm that evidence access is a prerequisite, not a guarantee. One case achieved accountability \emph{without} evidence access. In \emph{California EDD fraud} (2021), mass lockouts were reversed after public outcry and legislative pressure rather than evidentiary contestation. The system was abandoned wholesale through political channels, suggesting that when evidentiary contestation is impossible, affected parties are forced into expensive and uncertain collective action pathways.

\textbf{Sensitivity and robustness.} The 30\% ongoing rate creates uncertainty. Under pessimistic assumptions where ongoing access cases fail at 75\%, the access success rate drops to approximately 75\%, but the access/no-access differential remains large. Robustness checks (Appendix~\ref{appendix:supplementary}, Table~\ref{tab:robustness}) confirm the differential exceeds 75 percentage points across all specifications, including when excluding platform liability (99\% vs. 8\%), excluding criminal justice (92\% vs. 8\%), and examining pre-2020 cases only (84\% vs. 0\%).

\textbf{Platform liability and doctrinal immunity.} Six of the seven failures with evidence access are platform liability cases where Section 230 immunity barred liability despite full algorithmic scrutiny (\emph{Gonzalez v. Google}, \emph{Twitter v. Taamneh}, \emph{Force v. Facebook}, \emph{Dyroff v. Ultimate Software}, \emph{M.P. v. Meta}, \emph{Herrick v. Grindr}). A skeptic might argue that these undermine our thesis, but they demonstrate exactly what it predicts. Evidence access is required but insufficient alone. The plaintiffs in \emph{Gonzalez} and \emph{Force} examined recommendation algorithms in detail and demonstrated how those algorithms amplified harmful content. They lost because Section 230 immunizes such claims, not because they lacked evidence. This is precisely the outcome evidentiary rights would produce, with contestation reaching the merits rather than foundering on procedural obstacles. The remaining failure (\emph{Belgium Deliveroo}) involved an adverse determination on employment status, a merits loss rather than procedural foreclosure.

\textbf{Resource-level patterns.} High-resource litigants (government agencies, public interest organizations) obtained evidence access in 100\% of resolved cases (85/85), leaving no comparison group. Among lower-resource litigants where both access conditions exist, the pattern holds. Private counsel cases with access succeeded at 100\% (17/17) excluding platform liability, versus 12\% (1/8) without access. The absence of no-access cases among high-resource litigants itself demonstrates that sophisticated actors recognize evidence access as essential and obtain it before proceeding.

\subsection{Testing the Core Claim: Explanation Quality vs. Evidence Access}

Our central argument is that explanation alone does not enable accountability, and that evidence access is what matters. To test this, we coded explanation quality independently for cases where organizational communications were available. If explanation were sufficient, we would expect high-explanation cases to succeed regardless of evidence access.

\begin{table}[!ht]
\caption{The explanation-without-evidence test (resolved cases with explanation coding, n=121)\protect\footnotemark}
\label{tab:explanation-evidence}
\small
\centering
\renewcommand{\arraystretch}{1.15}
\begin{tabular}{@{}lccccc@{}}
\toprule
 & \multicolumn{2}{c}{\textbf{With Evidence Access}} & \phantom{ab} & \multicolumn{2}{c}{\textbf{No Evidence Access}} \\
\cmidrule{2-3} \cmidrule{5-6}
\textbf{Explanation} & Achieved & Denied & & Achieved & Denied \\
\midrule
High & 34 & 2 & & 0 & 4 \\
Partial & 34 & 3 & & 1 & 4 \\
None & 33 & 2 & & 0 & 4 \\
\midrule
\textbf{Total} & 101 & 7 & & 1 & 12 \\
\bottomrule
\end{tabular}
\end{table}
\footnotetext{Cell sizes in the ``No Evidence Access'' columns are small (n=3--5 per cell); we interpret these differences as suggestive of a pattern rather than statistically conclusive. The key finding is the contrast between columns: access cases succeed at 94\% regardless of explanation quality, while no-access cases fail at high rates across all explanation levels.}

Table~\ref{tab:explanation-evidence} reveals a suggestive pattern. Among cases \emph{with} evidence access, explanation quality shows no association with success, with rates of 94\% (high), 92\% (partial), and 94\% (none). Among cases \emph{without} evidence access, rates are uniformly low at 0\% (high), 20\% (partial), and 0\% (none). This is a striking result. Four cases received high-quality explanations with detailed factor disclosure but lacked evidence access, and all four failed. The robust finding is the column-level contrast. Cases with access succeed at 93\% regardless of explanation quality, while cases without access fail at 92\% across all explanation levels (Fisher's exact, $p < 0.001$). In other words, what the organization \emph{tells} the affected party matters far less than whether the affected party can independently \emph{verify} the decision. Full analysis appears in Appendix~\ref{appendix:supplementary}.
\subsection{What Evidence Access Provides}

Across successful cases, evidence access enabled three distinct contestation mechanisms (Table~\ref{tab:mechanisms}). \emph{Verification} allows affected parties to confirm that inputs were processed accurately. \emph{Comparison} allows them to compare their treatment to similar cases or aggregate patterns. \emph{Testing} allows them to probe system behavior by submitting modified inputs and observing results. These are precisely what current explanation requirements fail to provide. A person told ``your risk score was 0.73'' cannot verify the calculation, compare to similar cases, or test alternatives. GDPR Article 22's ``meaningful information,'' FCRA's factor disclosure, and platform appeal rationales all provide explanation without evidence access.

\begin{table}[!ht]
\caption{Fair contestation mechanisms enabled by evidence access}
\label{tab:mechanisms}
\small
\centering
\begin{tabular}{@{}lll@{}}
\toprule
\textbf{Mechanism} & \textbf{What it enables} & \textbf{Requires} \\
\midrule
Verification & Confirm inputs/processing accurate & Access to inputs, outputs, or logic \\
Comparison & Fair comparison to similar cases & Comparable cases or aggregate patterns \\
Testing & Probe behavior with variations & Submit modified inputs, observe results \\
\bottomrule
\end{tabular}
\end{table}

\textbf{Why the pattern holds across domains.} Despite substantial doctrinal differences, the evidence-access pattern is consistent across domains (Figure~\ref{fig:domain-outcomes}). This suggests evidentiary access functions as a \emph{meta-procedural invariant}. Regardless of whether the claim is employment discrimination or benefits termination, affected parties must verify inputs, compare treatment, and test variations. The specific legal framework varies, but the evidentiary structure of contestation does not. Different system types may require different evidentiary approaches, and sector-specific calibration is an important direction for future work. The absence of correlation between explanation quality and case success (Table~\ref{tab:explanation-evidence}) reinforces this point. In cases with evidence access, the explanation becomes redundant because the evidence itself reveals system behavior. In cases without access, even high-quality explanations cannot substitute for the ability to verify, compare, and test.

\section{Why Existing Frameworks Fail}
\label{sec:existing-frameworks}

We examine five major accountability frameworks, showing that each addresses Type 1 (opacity) while leaving Type 2 (evidentiary access) unaddressed.

\begin{enumerate}
    \item \textbf{GDPR Article 22} grants a right to ``contest'' automated decisions~\cite{gdpr2016,selbst2017meaningful}, but in practice yields only explanations. The CJEU's \emph{SCHUFA} decision (C-634/21, 2023) strengthened individual rights by holding that automated scoring constitutes a ``decision'' under Article 22, yet did not address whether affected parties can \emph{test} the system. Evidentiary access would add verification that inputs were accurately processed, comparison to similarly situated individuals, and testing of whether corrections to disputed data change the outcome.
    \item \textbf{FCRA adverse action} requires that consumers denied credit receive specific factors (e.g., ``high debt-to-income ratio'')~\cite{fcra1970} and is one of the most developed explanation regimes globally, yet it provides no verification capacity. A consumer cannot confirm the ratio was calculated correctly, test whether correcting a disputed debt would change the decision, or compare her treatment to similar applicants. Evidentiary access would add quantitative factor contributions, threshold disclosure, and a mechanism to test whether disputed items actually change outcomes.
    \item \textbf{Platform appeals} on major platforms provide content removal explanations citing community guidelines, but users cannot verify what triggered classification or whether rephrasing would change the result. Review panels exercise unbounded discretion with no published standards or precedent systems~\cite{mulligan2019shaping}. Evidentiary access would add identification of triggering features, testing of alternative formulations, and access to comparable cases.
    \item \textbf{Audit regimes} such as the EU AI Act and NYC Local Law 144 provide population-level accountability~\cite{euaiact2024,kroll2017accountable,raji2020closing} but do not help individuals contest specific decisions. An applicant cannot use an aggregate finding of ``3\% disparate impact'' to contest her personal denial. Individual-level evidentiary access would complement rather than replace system-level audits.
    \item \textbf{Anti-discrimination and human rights statutes} such as Title VII, the Equality Act 2010, and EU anti-discrimination directives provide substantive rights but no algorithmic-specific evidentiary mechanisms. Plaintiffs must prove discrimination through traditional discovery after filing a lawsuit, which is costly and time-consuming. Evidentiary rights would provide a pre-litigation pathway for generating evidence of algorithmic sensitivity to protected characteristics.
\end{enumerate}

This concentration on Type 1 persists for structural reasons. \emph{Technical tractability} means that XAI offers well-defined problems while evidentiary access requires institutional design~\cite{lipton2018mythos}. \emph{Organizational convenience} means that static disclosures require no infrastructure. \emph{Regulatory path dependence} means that frameworks translate ``explain'' but lack concepts for ``testable access.'' And \emph{legitimacy dynamics} means that explanation may perform accountability without actually enabling it~\cite{selbst2019fairness,edelman2016working}. We do not dismiss explanation but argue it cannot substitute for evidentiary access when the goal is contestation.
\section{Evidentiary Rights: Counterfactual Interrogation}
\label{sec:evidentiary-rights}

Our legal analysis identified three mechanisms enabling successful contestation: verification, comparison, and testing. We propose \emph{counterfactual interrogation rights} to provide these mechanisms without disclosing proprietary internals.

\textbf{Distinction from counterfactual explanations.} Counterfactual explanations~\cite{wachter2017counterfactual} are \emph{organization-generated}. For example, ``you would have been approved if your income were \$5,000 higher.'' Organizations control which features to vary, enabling strategic selection. Counterfactual interrogation is \emph{affected-party-controlled}, meaning individuals submit their own variations and observe responses, testing factors they suspect are problematic. The former tells you what organizations want you to know, while the latter lets you discover what you want to know.

Under counterfactual interrogation, affected parties receive a sandboxed testing interface to submit modified inputs and observe whether assessments change. Significant divergence between semantically similar inputs (e.g., ``anxiety'' vs. ``anxiety symptoms'' producing different outcomes) creates documented grounds for appeal, providing evidence that the system responds to surface features rather than meaning. The interface returns scores, confidence intervals, and percentiles but does \emph{not} return model internals or training data.

The mechanism is triggered by material algorithmic involvement in an adverse decision, and affected parties may request access within 30 days. Organizations must test against the model version that made the decision (analogous to the spoliation doctrine, where loss of relevant evidence creates an adverse inference against the party responsible for preserving it). Similarity is defined through regulator-mandated perturbation classes, which are domain-specific categories of semantically equivalent variations (e.g., name formatting, terminology updates, framing alternatives) that should not affect legitimate assessments (detailed specification in Appendix~\ref{appendix:mechanism}). Because access is limited, rate-bounded, and tied to adverse decisions rather than open querying, counterfactual interrogation avoids the boundary-mapping risks associated with unrestricted model access.

\textbf{Divergence thresholds.} We propose four concrete criteria, any of which creates grounds for appeal:
\begin{enumerate}
    \item \emph{Outcome-crossing}. A perturbation moves the assessment from reject to accept (or vice versa). For example, changing graduation year from 1991 to 2011 flips a hiring score from 0.42 to 0.71, crossing the 0.60 threshold.
    \item \emph{Percentile shift $>$15 points}. The perturbation moves the applicant's ranking by more than 15 percentile points. For example, a name change moves an applicant from 23rd to 41st percentile.
    \item \emph{Threshold-proximate shift $>$5 points}. For near-threshold cases (within 10 percentile points), a shift of $>$5 points triggers review. For example, an applicant at the 56th percentile (threshold 60th) drops to 48th.
    \item \emph{Pattern-consistent effects}. The perturbation produces effects consistent with known discrimination patterns (e.g., age-related terms systematically reduce scores), requiring $\geq$3 consistent perturbations.
\end{enumerate}

These thresholds are policy-tunable parameters that balance sensitivity against false positives, and their function is to demonstrate feasibility rather than prescribe universal values (justification in Appendix~\ref{appendix:mechanism}). Percentile-based thresholds (criteria 2 and 3) apply most directly to systems producing continuous scores or rankings. For binary or categorical decisions, outcome-crossing (criterion 1) and pattern-consistent effects (criterion 4) are more appropriate. Domain regulators should specify which criteria apply to which system types. Divergence creates grounds for appeal, not automatic reversal, and organizations must justify different treatment. This applies whether the algorithmic score is the final decision or merely advisory input to a human decisionmaker. If a perturbation changes the score substantially but the organization claims a human independently reached the same conclusion, the divergence still warrants justification. Either the score influenced the human, requiring explanation of its sensitivity, or it did not, raising questions about why the system is deployed at all. Importantly, semantic equivalence is itself contested. Whether ``25 years experience'' is equivalent to ``extensive experience'' depends on whether precision is legitimately job-relevant. The mechanism does not resolve this algorithmically but instead surfaces the question for adjudicator judgment at both stages (Appendix~\ref{appendix:mechanism}).

\begin{figure}[ht]
\centering
\scalebox{0.8}{%
\begin{tikzpicture}[
    node distance=0.6cm,
    process/.style={rectangle, rounded corners=3pt, draw=cbblue!70, fill=cbblue!8, line width=0.6pt, minimum width=4.2cm, minimum height=0.75cm, align=center, font=\footnotesize},
    decision/.style={diamond, draw=cborange!80, fill=cborange!10, line width=0.7pt, minimum width=2.2cm, minimum height=1.2cm, align=center, font=\footnotesize, aspect=2.5},
    success/.style={rectangle, rounded corners=3pt, draw=cbblue, fill=cbblue!15, line width=0.7pt, minimum width=4.2cm, minimum height=0.75cm, align=center, font=\footnotesize},
    neutral/.style={rectangle, rounded corners=3pt, draw=black!30, fill=gray!5, line width=0.5pt, minimum width=3.2cm, minimum height=0.7cm, align=center, font=\footnotesize},
    arrow/.style={-{Stealth[length=2.5mm, width=2mm]}, line width=0.6pt, black!50},
    successarrow/.style={-{Stealth[length=2.5mm, width=2mm]}, line width=0.7pt, cbblue!80},
]

\node[process] (step1) {\textbf{1.} Adverse Decision Received};
\node[process, below=of step1] (step2) {\textbf{2.} Portal Access Granted};
\node[process, below=of step2] (step3) {\textbf{3.} Submit Test Perturbations};
\node[process, below=of step3] (step4) {\textbf{4.} Observe Output Changes};
\node[decision, below=0.7cm of step4] (decide) {Divergence?};
\node[success, below left=0.8cm and 0.3cm of decide] (step5) {\textbf{5.} File Appeal with Evidence};
\node[neutral, below right=0.8cm and 0.3cm of decide] (nogrounds) {No grounds found};
\node[success, below=0.6cm of step5] (step6) {\textbf{6.} Independent Adjudication};

\draw[arrow] (step1) -- (step2);
\draw[arrow] (step2) -- (step3);
\draw[arrow] (step3) -- (step4);
\draw[arrow] (step4) -- (decide);

\draw[successarrow] (decide.south west) -- ++(0,-0.25) -| node[pos=0.25, above, font=\footnotesize\bfseries, text=cbblue] {Yes} (step5.north);
\draw[arrow] (decide.south east) -- ++(0,-0.25) -| node[pos=0.25, above, font=\footnotesize, text=black!50] {No} (nogrounds.north);
\draw[successarrow] (step5) -- (step6);

\draw[arrow, densely dotted] (nogrounds.east) -- ++(0.5,0) |- node[pos=0.25, right, font=\scriptsize, text=black!45] {Retry} (step3.east);

\end{tikzpicture}}%
\caption{Counterfactual interrogation process flow. Steps 1--4 are sequential: adverse decision received (1), portal access granted (2), perturbations submitted (3), output changes observed (4). A divergence assessment (diamond) then branches: significant divergence leads to filing an appeal with evidence (5) and independent adjudication (6); no divergence allows retesting with different factors.}
\label{fig:process-flow}
\end{figure}
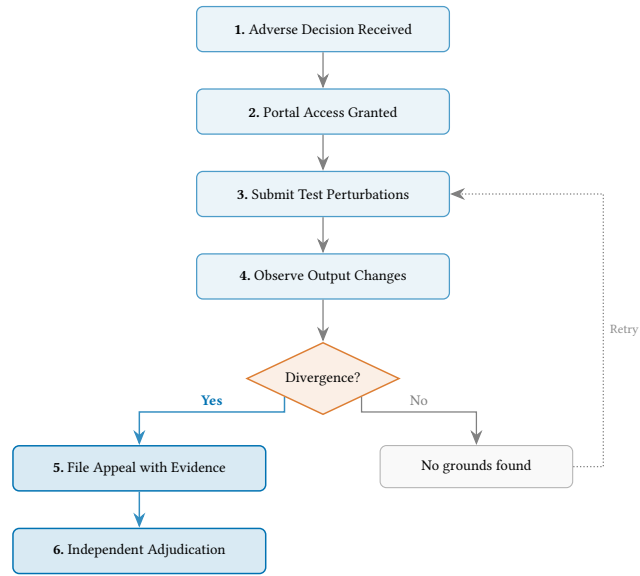

To illustrate how this works in practice, consider Maria, 58, who is rejected by an AI hiring screen with a score of 0.42. She tests perturbations and finds that changing her graduation year from 1991 to 2011 raises the score to 0.71, while a name change raises it to 0.58. This documented divergence creates evidence for appeal, and the employer must explain why these variations reflect legitimate distinctions. A full worked example appears in Appendix~\ref{appendix:example}.

Counterfactual interrogation differs fundamentally from traditional litigation discovery (Table~\ref{tab:discovery-compare}). Discovery requires filing a lawsuit first, takes months to years, requires attorneys, and produces documents that must be interpreted. Counterfactual interrogation triggers automatically upon adverse decisions, completes in days, requires no counsel, and produces behavioral evidence directly relevant to the contested decision. The affected party controls testing rather than engaging in adversarial negotiation over document production. Enforcement relies on mandatory logging with 3-year retention, burden-shifting for missing evidence, default presumptions against non-materiality claims, and non-compliance penalties.

\begin{table}[!ht]
\caption{Counterfactual interrogation vs. litigation discovery}
\label{tab:discovery-compare}
\small
\centering
\begin{tabular}{@{}l@{\hspace{3.5em}}l@{\hspace{3.5em}}l@{}}
\toprule
\textbf{Dimension} & \textbf{Litigation Discovery} & \textbf{Counterfactual Interrogation} \\
\midrule
Trigger & Lawsuit filed & Adverse decision \\
Timeline & Months--years & Days--weeks \\
Cost & High (attorneys required) & Low (self-service) \\
Control & Adversarial negotiation & Affected-party directed \\
Output & Documents to interpret & Behavioral divergence evidence \\
Counsel required & Effectively yes & No \\
\bottomrule
\end{tabular}
\end{table}

\textbf{Gaming objection.} A full analysis appears in Appendix~\ref{appendix:gaming}. \emph{Gaming by affected parties} is manageable for three reasons. First, access is to assessments, not operational outputs. Second, rate limits (50 queries) prevent boundary-mapping. Third, fair housing testing has operated since 1968 despite theoretical gaming concerns~\cite{selmi2006testing}. Although algorithmic systems produce continuous scores over high-dimensional spaces unlike binary fair housing tests, this distinction \emph{favors} our proposal because sensitivity detection is equally feasible for both, while adversarial extraction is \emph{harder} for high-dimensional systems. Direct access is appropriate for most domains, while fraud detection and national security warrant mediated access via licensed representatives. The 50-query limit balances anti-gaming with legitimate use, as default perturbation suites cover standard protected characteristics in 10--15 queries and remaining queries allow hypothesis-driven testing of case-specific concerns. This limit is a regulatory parameter, adjustable based on empirical evidence. \emph{Gaming by organizations} poses a distinct risk. Companies could modify outputs over common perturbation suites to suppress divergence during testing while maintaining discriminatory behavior on live inputs. Regulatory countermeasures include unannounced audit testing by agencies using non-standard perturbations, mandatory logging of whether test-mode processing differs from production, and penalties for differential treatment of testing versus operational queries (analogous to emissions testing fraud).

\textbf{Complementary requirements.} The remaining failure types also require attention. For Type 3 (forums), existing agencies suffice for regulated domains while platforms need independent adjudication. For Type 4 (standards), divergence creates a rebuttable presumption and organizations must articulate legitimate reasons. For Type 5 (remedies), individual relief, system modification, and penalties for obstruction are needed.

\section{Implementation Pathways}
\label{sec:regulatory-comparison}

Counterfactual interrogation is designed primarily for settings where no lawsuit is involved, including administrative appeals, workplace disputes, and platform contexts. The affected party tests the system, generates evidence, and presents findings through existing channels. If the organization justifies divergence, the matter resolves without litigation. Otherwise, documented evidence supports formal proceedings.

Evidentiary rights need not require new legislation. GDPR Article 22 could be reinterpreted so that ``meaningful information'' includes verification ability and testing access. FCRA adverse action could require quantitative factor contributions and threshold disclosure. ADA reasonable accommodation could encompass alternative assessment methods and human review with evidence access. Beyond reinterpretation, new mechanisms could also help. EU AI Act conformity assessments could require testing interface capability, and sector-specific mandates could extend existing authorities such as EEOC for employment testing, CFPB for credit counterfactual disclosure, and HUD for algorithmic fair housing testing. Platform accountability legislation could require specific content identification, appeal rights with evidence access, and independent adjudication for consequential decisions. For regulated domains, existing agencies can integrate testing into current enforcement. For unregulated domains, private rights of action and state attorney general authority provide alternatives. Contractual flow-down requirements prevent avoidance through outsourcing to opaque third-party systems.

We propose graduated implementation across three tiers. Tier 1 domains (criminal justice, healthcare, employment, housing, and credit) would receive full counterfactual access. Tier 2 domains (content moderation, insurance, and education) would receive limited or representative-mediated access. Tier 3 domains (recommendations and advertising) would receive aggregate transparency only. Tier 3's flexibility reflects that recommendations and advertising produce lower-stakes, more easily reversible effects. Aggregate transparency provides proportionate accountability without imposing individual testing costs on billions of daily interactions. The scalability concern assumes every adverse decision triggers contestation, but analogous regimes suggest otherwise. FCRA dispute rights (since 1970) generate approximately 8 million disputes annually among 200 million credit-active consumers (4\% rate), and GDPR subject access requests generate 2--5 requests per 1000 data subjects annually. We anticipate similar patterns. Testing interfaces resemble existing A/B testing systems, and version retention uses standard ML versioning (MLflow, DVC). Based on FCRA dispute handling costs (\$5--10 per dispute) and GDPR response costs (\texteuro15--50 per request), we estimate testing infrastructure at single-digit percentages of deployment.

\textbf{Low-resource populations.} Evidentiary rights risk reproducing resource asymmetries. The cognitive burden of hypothesizing problematic factors and constructing perturbations may require professional assistance, as fair housing testing is conducted by trained professionals rather than individual seekers. We address this through five mechanisms (Appendix~\ref{appendix:low-resource}), including plain-language reports, publicly funded testing services, default perturbation suites, accessibility mandates, and collective action integration. These reduce but do not eliminate the sophistication barrier. The practical burden on affected parties varies by tier. For Tier 1 decisions, default perturbation suites run automatically upon request, plain-language reports summarize results, and publicly funded testing services provide professional assistance for complex cases. The minimum burden is to request testing and review the report, comparable to requesting a free credit report.

\textbf{Case study.} Consider \emph{Louis v. SafeRent} (2024), where a Section 8 voucher holder was denied housing based on SafeRent's algorithmic tenant screening. The plaintiff obtained access through litigation discovery, revealing that the algorithm penalized prior eviction records without distinguishing retaliatory evictions from fault-based ones, producing discriminatory effects on Black applicants. Under evidentiary rights, the applicant would instead request testing access within 30 days of denial. Perturbations with the eviction record removed, different eviction types, and demographic variations would reveal that eviction history produced a 35-percentile shift while the nature of the eviction was ignored. The report would be submitted to HUD, and SafeRent would need to justify why it treats retaliatory and fault-based evictions identically. The process would take days, not years of litigation, without requiring counsel.

\section{Limitations}
\label{sec:limitations}

Evidentiary rights address a critical gap in algorithmic accountability, but they are not a complete solution. Several important limitations should be noted.

\begin{itemize}
    \item \textbf{Resource and capacity constraints.} Formal rights without practical capacity produce paper accountability. Professional intermediaries may be necessary for effective contestation, limiting the mechanism's democratic potential, much as the right to counsel means little without funded public defenders.
    \item \textbf{Ritualistic compliance.} Power~\cite{power1997audit} and Edelman~\cite{edelman2016working} warn that formal mechanisms can legitimate existing practices without changing them. Counterfactual interrogation is more resistant to co-optation than static disclosure because it generates falsifiable evidence that regulators can independently verify, but no mechanism is immune to strategic compliance.
    \item \textbf{Scope.} Evidentiary rights address individual contestation only, not systemic harms, anticipatory governance, or collective action problems. Individual and systemic mechanisms are complements rather than substitutes. Individual evidence can trigger systemic investigation, and systemic findings can support individual claims.
    \item \textbf{Mediated access complexity.} For high-risk domains such as fraud detection and national security, we propose mediated access through licensed representatives, regulatory testing, or escrowed disclosure. These mechanisms add substantial implementation complexity and require institutional development beyond our scope.
    \item \textbf{Empirical limitations.} Our analysis shows correlation, not causation, and evidence access may proxy for case strength. Additionally, 30\% of cases remain ongoing, and platform liability cases show that access does not guarantee success.
    \item \textbf{International case heterogeneity.} Our 36 international cases span 19 jurisdictions with different legal frameworks. We cannot pool these for valid inference but include them to suggest the evidentiary barrier is not US-specific. Results appear separately in Appendix~\ref{appendix:cases}.
    \item \textbf{Research agenda for Types 3--5.} This paper addresses Type 2 only. Types 3--5 require separate research, and detailed questions appear in Appendix~\ref{appendix:future-work}.
\end{itemize}

\section{Conclusion}
\label{sec:conclusion}

Algorithmic accountability scholarship has achieved remarkable progress on explanation, yet affected parties remain unable to meaningfully contest decisions. Explanation and contestation are distinct: explanation tells you why; contestation enables you to challenge whether the decision was correct. Our analysis of 168 litigated cases finds that evidentiary access, not explanation quality, is strongly associated with accountability success. This association reflects selection effects we cannot fully resolve, but the diagnostic pattern is clear: among observable cases, contestation without evidence access fails at the procedural stage before reaching substantive evaluation. The prescription is evidentiary rights: capacity to verify inputs were processed correctly, compare treatment to similar cases, and test whether variations would change outcomes. Counterfactual interrogation provides this without disclosing proprietary internals.
Evidentiary rights do not replace explanation requirements, audit regimes, or impact assessments; they fill the gap between understanding a decision and challenging it. The path forward requires reinterpretation of existing frameworks (GDPR, FCRA, civil rights law) and new mechanisms (testing interfaces, sector mandates). Our loan applicant who understands her denial but cannot challenge it would gain capacity to verify her credit history was accurately assessed, test whether corrections would change the outcome, and compare her treatment to similar applicants. She would have not just an explanation but evidence. Evidence is what accountability requires.

\section*{Generative AI Usage Statement}
The authors used generative AI tools (Claude) for editing assistance, including grammar checking, prose refinement, and formatting suggestions. All substantive content, analysis, arguments, and conclusions were developed by the human authors. The legal case database was compiled through manual research and coding by the authors. AI-generated text was not used for the core intellectual contributions of this work.

\bibliographystyle{ACM-Reference-Format}
\bibliography{references}

\appendix

\section{Research Agenda for Types 3--5}
\label{appendix:future-work}

\emph{Type 3 (Forum).} What institutional forms enable effective algorithmic adjudication? Should forums be domain-specific or consolidated? How should platform-internal appeals relate to external review? What procedural rules govern disputes (discovery scope, burden allocation, expert testimony)? Comparative study of existing forums could identify design principles.

\emph{Type 4 (Standards).} What substantive standards should adjudicators apply? When does divergence indicate discrimination versus legitimate differentiation? How should ``legitimate business necessity'' defenses apply? Our framework proposes burden-shifting, but the content of those burdens requires normative and doctrinal analysis beyond our empirical scope.

\emph{Type 5 (Remedies).} What remedies follow successful contestation? Individual relief addresses case-specific harm but may not deter systemic problems. System modification addresses root causes but raises governance questions. Penalties incentivize compliance but may be captured or evaded.

\section{Implementation for Low-Resource Populations}
\label{appendix:low-resource}

\textbf{Plain-language divergence reports.} Organizations must provide mandatory divergence summaries in accessible language when counterfactual testing reveals significant differences. Rather than raw divergence statistics, reports must state: ``When we tested your application with [change], the outcome changed from [denied] to [approved]. This suggests that [factor] may have affected your decision.'' Regulatory agencies (CFPB, EEOC) would specify templates and reading-level requirements (8th grade or below), analogous to existing plain-language mandates for credit disclosures and employment notices.

\textbf{Publicly funded testing services.} Just as public defenders provide legal representation, publicly funded algorithmic testing services can provide technical capacity: (1) \emph{Self-service portals} with pre-configured perturbation suites and automated divergence flagging; (2) \emph{Community navigator programs} where trained staff at libraries and workforce centers assist with testing; (3) \emph{Technical assistance grants} funding civil society organizations to conduct testing on behalf of communities.

\textbf{Default perturbation suite governance.} Default perturbation suites must be designed through multi-stakeholder processes: regulatory agencies specify mandatory perturbation categories for their domains; affected-party representatives participate through notice-and-comment rulemaking; civil society organizations can petition to add categories based on documented concerns.

\textbf{Interface accessibility mandates.} Testing interfaces must meet: WCAG 2.1 AA compliance; multilingual support for languages representing $>$5\% of affected population; mobile-accessible design; offline capability; maximum 48-hour response times. Regulatory agencies should conduct mystery-shopper audits to identify evasive compliance.

\textbf{Collective action integration.} The framework should facilitate: pattern sharing across testers to identify systematic divergences; class action triggers when testing reveals consistent patterns; regulatory referral mechanisms when individual testing generates evidence of population-level discrimination.

\textbf{Addressing interpretive capacity.} We propose: mandatory ``next steps'' guidance; agency hotlines for divergence questions; integration with existing complaint infrastructure (results automatically attachable to EEOC, CFPB, or HUD complaints).

\textbf{The role of professional intermediaries.} Effective contestation may require professional assistance for most affected parties. Fair housing testing is conducted by trained testers working for civil rights organizations, not individual seekers. Counterfactual interrogation may function most effectively when conducted by civil society organizations, legal aid attorneys, union representatives, or publicly funded testing services. This constrains the mechanism's democratic potential but does not render it valueless; fair housing testing has produced significant accountability despite professional mediation.

{\color{blue}
\section{Case Identification Flow}
\label{appendix:prisma}
}

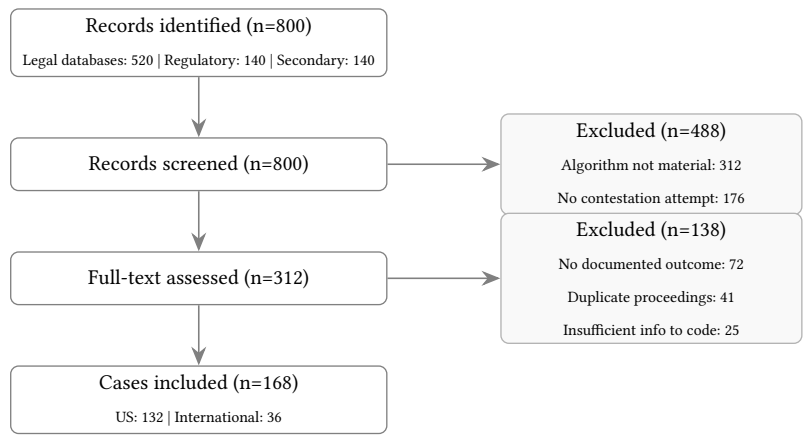
\begin{figure}[H]
\centering
\begin{tikzpicture}[
    box/.style={rectangle, rounded corners=3pt, draw=black!50, fill=white, minimum width=5cm, minimum height=0.7cm, align=center, font=\small, line width=0.5pt},
    excl/.style={rectangle, rounded corners=3pt, draw=black!30, fill=gray!5, minimum width=4cm, minimum height=0.7cm, align=center, font=\small, line width=0.5pt},
    arrow/.style={-{Stealth[length=2.5mm, width=2mm]}, line width=0.6pt, black!50},
    node distance=0.8cm,
]
\node[box] (id) {Records identified (n=800)\\{\scriptsize Legal databases: 520 | Regulatory: 140 | Secondary: 140}};
\node[box, below=of id] (screen) {Records screened (n=800)};
\node[excl, right=1.5cm of screen] (excl1) {Excluded (n=488)\\{\scriptsize Algorithm not material: 312}\\{\scriptsize No contestation attempt: 176}};
\node[box, below=of screen] (elig) {Full-text assessed (n=312)};
\node[excl, right=1.5cm of elig] (excl2) {Excluded (n=138)\\{\scriptsize No documented outcome: 72}\\{\scriptsize Duplicate proceedings: 41}\\{\scriptsize Insufficient info to code: 25}};
\node[box, below=of elig] (incl) {Cases included (n=168)\\{\scriptsize US: 132 | International: 36}};

\draw[arrow] (id) -- (screen);
\draw[arrow] (screen) -- (elig);
\draw[arrow] (elig) -- (incl);
\draw[arrow] (screen) -- (excl1);
\draw[arrow] (elig) -- (excl2);
\end{tikzpicture}
\caption{PRISMA-style case identification flow diagram. Records were identified through legal database searches (Westlaw, LexisNexis), regulatory enforcement databases (EEOC, CFPB, FTC, DOJ), and secondary sources (academic surveys, investigative journalism, advocacy databases). International cases were drawn from European court databases, data protection authority decisions, and comparative law sources.}
\label{fig:prisma}
\end{figure}

\section{Supplementary Statistical Analysis}
\label{appendix:supplementary}

\subsection{Controlling for Alternative Explanations}

Table~\ref{tab:legal-resources-app} shows evidence access distribution by legal resources.

\begin{table}[!ht]
\caption{Evidence access and legal resources (resolved cases, n=121)}
\label{tab:legal-resources-app}
\small
\begin{tabular}{@{}lccc|c@{}}
\toprule
\textbf{Legal Resources} & \textbf{Full} & \textbf{Partial} & \textbf{None} & \textbf{Total} \\
\midrule
High (government) & 16 & 18 & 0 & 34 \\
Medium (public interest/union/academic) & 22 & 29 & 4 & 55 \\
Low (private counsel/pro se) & 13 & 10 & 9 & 32 \\
\midrule
\textbf{Total} & 51 & 57 & 13 & 121 \\
\bottomrule
\end{tabular}
\end{table}

High-resource litigants (government agencies) obtained evidence access in 100\% of resolved cases (34/34), leaving no comparison group. Among medium-resource litigants (public interest organizations, unions, academics), success was 98\% (50/51) with access versus 0\% (0/4) without. Among low-resource litigants (private counsel, pro se), success was 74\% (17/23) with access versus 11\% (1/9) without; excluding platform liability cases (where Section 230 immunity caused failure despite access), the with-access rate rises to 100\% (17/17). Fisher's exact test confirms the access/no-access differential is significant ($p < 0.001$). The absence of no-access cases among high-resource litigants demonstrates that sophisticated actors recognize evidence access as essential.

\subsection{Access Level Outcomes}

Success rates differ by access level: Full access 88\% (45/51), Partial access 97\% (56/58), None 9\% (1/11). The lower Full access rate reflects platform liability cases where Section 230 immunity blocked liability despite complete algorithmic scrutiny (6 of 6 Full-access failures). Excluding platform liability, Full access success is 97\% (45/46).

\subsection{Robustness Checks}

We conducted robustness checks to assess whether results depend on specific domain inclusion or coding decisions.

\textbf{Excluding criminal justice.} Criminal justice cases (n=23 resolved) have distinct dynamics: constitutional protections, different discovery norms, and higher public scrutiny. Excluding this domain: with access 79/86 (92\%), without access 1/12 (8\%). Fisher's exact $p < 0.001$. Pattern holds.

\textbf{Excluding platform liability.} Platform liability cases (n=6 resolved) demonstrate that evidence access is necessary but not sufficient when doctrinal immunity applies. Excluding this domain: with access 101/102 (99\%), without access 1/13 (8\%). This provides the clearest test of whether evidence access is associated with success absent doctrinal barriers.

\textbf{Pre-2020 vs. 2020+.} Earlier cases (2012--2019, n=24): with access 16/19 (84\%), without access 0/5 (0\%). Recent cases (2020--2025, n=97): with access 85/89 (96\%), without access 1/8 (12\%). Pattern is stable across time periods, with higher success rates in recent cases potentially reflecting improved litigation strategies and precedent development.

\begin{table}[!ht]
\caption{Robustness checks summary (n=121 resolved)}
\label{tab:robustness}
\small
\begin{tabular}{@{}lcccc@{}}
\toprule
\textbf{Analysis} & \textbf{Access} & \textbf{No Access} & \textbf{Diff} & \textbf{$p$} \\
\midrule
Full sample (resolved) & 93\% (100/107) & 9\% (1/11) & 84pp & $<$.001 \\
Excl. platform liability & 99\% (101/102) & 8\% (1/13) & 91pp & $<$.001 \\
Excl. criminal justice & 92\% (79/86) & 8\% (1/12) & 84pp & $<$.001 \\
Pre-2020 cases only & 84\% (16/19) & 0\% (0/5) & 84pp & $<$.001 \\
\bottomrule
\end{tabular}
\end{table}

\section{Mechanism Design Details}
\label{appendix:mechanism}

\subsection{Threshold Justification}

The divergence thresholds derive from three independent considerations: effect size significance, noise tolerance, and administrative feasibility---not from analogies to aggregate statistical tests like the 4/5ths rule, which measures disparate impact across groups rather than individual-level sensitivity.

\textbf{Effect size rationale.} The 15-percentile threshold reflects a judgment about practical significance: a perturbation that moves an individual's ranking by 15+ percentile points represents a substantial change in competitive position. For context, in a pool of 100 applicants, this means moving from position 50 to position 35 or worse---a difference likely to affect selection outcomes in competitive processes. The threshold asks: ``Is this change large enough that the affected party's competitive position meaningfully differs?''

\textbf{Noise tolerance.} Algorithmic systems exhibit inherent variance from tokenization, embedding choices, and model stochasticity. Informal testing of commercial systems suggests semantically identical inputs typically produce score variations of 2--5 percentile points, though this estimate lacks peer-reviewed validation and likely varies by system. The 15-point threshold is designed to sit above plausible noise floors; actual thresholds should be calibrated empirically for specific domains through stakeholder input, not adopted from this illustrative specification.

\textbf{Administrative calibration.} Sensitivity analysis informs threshold selection: at 10-percentile thresholds, approximately 40\% of perturbations trigger review (high false-positive burden on adjudicators); at 20 percentiles, approximately 8\% trigger (potentially missing legitimate concerns). The 15-point threshold balances detection sensitivity against administrative capacity. Domain-specific calibration may adjust these baselines; employment decisions affecting livelihood may warrant tighter thresholds (10--12 points) than content moderation (18--20 points), reflecting different stakes and acceptable error rates.

\textbf{What thresholds do not establish.} These thresholds create \emph{grounds for appeal}, not determinations of discrimination. A 20-percentile shift from changing a graduation year suggests the system is age-sensitive; whether that sensitivity reflects unlawful discrimination or legitimate job-relatedness is determined through adjudication, not threshold-crossing alone.

\subsection{Discovery Comparison}

\begin{table}[!ht]
\caption{Counterfactual interrogation vs. litigation discovery}
\label{tab:discovery-app}
\small
\centering
\begin{tabular}{@{}lll@{}}
\toprule
\textbf{Dimension} & \textbf{Discovery} & \textbf{Counterfactual Interrogation} \\
\midrule
Trigger & Lawsuit filed & Adverse decision \\
Timeline & Months--years & Days--weeks \\
Cost & High (attorneys) & Low (self-service) \\
Control & Adversarial & Affected-party directed \\
Output & Documents & Divergence evidence \\
Counsel required & Effectively yes & No \\
\bottomrule
\end{tabular}
\end{table}

\subsection{Interface Specifications}

The testing interface returns: (1) assessment score for each input; (2) confidence interval; (3) percentile ranking. It does \emph{not} return model internals, training data, or other users' information.

\subsection{Perturbation Space Definition}

A valid perturbation must satisfy the \emph{semantic equivalence constraint}: the modified input should preserve the substantive meaning relevant to the legitimate assessment purpose while varying surface features that should be irrelevant. This is not a formal mathematical constraint but a normative judgment operationalized through domain-specific perturbation classes.

\textbf{Perturbation classes by domain:}
\begin{itemize}
    \item \emph{Employment}: Name variations (ethnic/gender markers), date reformatting (graduation year presentation), terminology updates (``J2EE'' vs. ``Java enterprise''), experience framing (``25 years'' vs. ``extensive''), institution naming (``State University'' vs. full name).
    \item \emph{Credit/lending}: Address formatting, name transliteration, employment description variations, income presentation (hourly vs. annual), date formats.
    \item \emph{Healthcare}: Symptom synonyms (``anxiety'' vs. ``nervousness''), temporal framing (``chronic'' vs. ``ongoing''), severity descriptions, medical terminology variations.
    \item \emph{Content moderation}: Register shifts (formal/informal), dialect variation, indirect vs. direct framing, hedging language, emotional valence markers.
\end{itemize}

\textbf{What perturbations test.} Each perturbation class targets a specific hypothesis: name variations test for demographic proxies; terminology variations test for recency bias; framing variations test for semantic robustness. The affected party's burden is to identify \emph{which} features they suspect are problematic and construct perturbations testing those hypotheses. The mechanism does not require exhaustive search---it requires targeted hypothesis testing.

\textbf{Semantic equivalence is contested, not given.} A critical limitation: contested cases likely involve precisely those inputs where semantic equivalence is disputed. Is ``25 years experience'' semantically equivalent to ``extensive experience''? The answer depends on whether length of experience is legitimately job-relevant---exactly what is being contested. We do not assume this problem away. Rather, the mechanism shifts burden: divergence creates a rebuttable presumption that the organization must address. If an employer argues that ``25 years'' genuinely differs from ``extensive'' because precision matters for the role, that argument is evaluated by an adjudicator. The perturbation does not automatically establish discrimination; it establishes that the system is sensitive to the variation, requiring justification. This means adjudicator judgment operates at two stages: (1) whether the perturbation represents a variation that should be outcome-irrelevant, and (2) whether the organization's justification for sensitivity is legitimate. The mechanism surfaces the contested question rather than resolving it algorithmically.

\textbf{Regulators specify minimum suites.} Domain regulators (EEOC for employment, CFPB for credit, HUD for housing) would specify minimum perturbation suites that all testing interfaces must support, ensuring baseline coverage of common discrimination vectors. These regulatory determinations would themselves be subject to notice-and-comment rulemaking, allowing stakeholders to contest whether specific perturbation classes represent legitimately equivalent variations. Affected parties may propose additional perturbations beyond the minimum suite, subject to adjudicator assessment of relevance.

\textbf{Governance example.} CFPB proposes credit perturbation classes: name transliterations, address formats, income presentation (hourly vs. annual). During notice-and-comment, lenders argue income presentation differences are substantive (hourly workers have different risk profiles) rather than stylistic. Final rule accepts name/address classes but requires case-by-case adjudication for income presentation variations, with the burden on lenders to justify differential treatment. Annual review adds new class (employment verification date variations) based on enforcement patterns. This illustrates that perturbation class governance is contestable, iterative, and subject to standard administrative law constraints.

\subsection{Version-Pinning Scenarios}

\emph{Same version available}: Test against decision-time version. \emph{Updated, original retained}: Test original; current-version differences may support defect claims. \emph{Original not retained}: Presumption favors affected party (spoliation analogy).

\section{Worked Example: Employment Screening}
\label{appendix:example}

Maria, 58, applies for senior engineering. Rejected with score 0.42 (threshold 0.60). Notification: ``Algorithmic screening contributed. Testing rights available for 30 days.''

\textbf{Testing}: (1) Graduation year 1991$\rightarrow$2011: score 0.42$\rightarrow$0.71. (2) Name ``Maria Gonzalez''$\rightarrow$``Michael Gordon'': 0.42$\rightarrow$0.58. (3) ``25 years enterprise''$\rightarrow$``extensive full-stack'': 0.42$\rightarrow$0.64. (4) ``J2EE, SOAP''$\rightarrow$``Spring Boot, REST'': 0.42$\rightarrow$0.69.

\textbf{Divergence report}: ``Significant divergence detected. Graduation year [0.29], name [0.16], framing [0.22], terminology [0.27]. Assessment may be sensitive to factors unrelated to qualifications.''

\textbf{Appeal}: Maria documents that age/name proxies affected her score. Employer must explain why variations reflect legitimate distinctions. If inadequate (``trained on historical data, we don't know why''), Maria receives relief.

\section{Gaming Objection Details}
\label{appendix:gaming}

The primary objection to counterfactual interrogation rights is that they enable gaming: affected parties could learn to manipulate inputs to obtain favorable outcomes without genuine qualification changes. We address this objection through threat modeling, empirical precedent, and design mitigations.

\subsection{Threat Model}

\begin{table}[!ht]
\caption{Threat model for counterfactual interrogation}
\label{tab:threat-app}
\small
\centering
\begin{tabular}{@{}llll@{}}
\toprule
\textbf{Actor} & \textbf{Capability} & \textbf{Risk} & \textbf{Mitigation} \\
\midrule
Legitimate party & 50 queries & Low & Intended use \\
Sophisticated & Repeated apps & Medium & Cross-app limits \\
Coordinated & Pooled queries & Med-High & Pattern detection \\
Resourced adversary & Distributed & High & Escrowed reps \\
\bottomrule
\end{tabular}
\end{table}

\textbf{Query budget analysis.} With 50 queries and typical input dimensionality (10-50 features), an adversary can probe $<1\%$ of the decision space. Binary search on a single feature requires $\log_2(\text{range})$ queries; probing 10 features independently exhausts the budget without learning interaction effects. Decision boundaries for high-dimensional systems cannot be meaningfully reconstructed from 50 queries.

\textbf{Rate limiting.} Per-applicant limits (50 queries/decision) prevent exhaustive probing. Cross-application tracking (10 queries/month across all applications) limits repeated probing. Anomaly detection flags coordinated patterns (e.g., 100 applicants submitting identical perturbation sequences).

\textbf{Sufficiency for legitimate contestation.} Fifty queries is sufficient for legitimate use because contestation requires sensitivity detection, not exhaustive mapping. Default perturbation suites test standard protected characteristics (race, gender, age, disability) in approximately 10-15 queries. Remaining queries allow affected parties to test case-specific hypotheses (e.g., ``did my gap year matter?''). Legitimate contestation rarely requires testing more than 5-10 factors; each factor requires only 2-5 perturbations to detect sensitivity. If empirical evidence suggests legitimate users routinely exhaust budgets without obtaining useful evidence, the limit should be increased; if gaming emerges, it should be decreased or mediated access expanded. The 50-query figure is a starting point for regulatory calibration, not a fixed parameter.

\subsection{Empirical Precedent: Fair Housing Testing}

Fair housing testing has operated since 1968 under the Fair Housing Act. Testers submit paired applications differing only on protected characteristics to detect discrimination. The predicted gaming catastrophe never materialized.

\textbf{The binary vs.\ high-dimensional distinction.} Fair housing tests binary decisions; algorithmic systems produce continuous scores over high-dimensional spaces. Does this difference matter? For \emph{adversarial extraction}, yes: reconstructing high-dimensional functions requires exponentially more queries. But counterfactual interrogation is sensitivity detection, not extraction. The question ``does the system respond differently to equivalent inputs?'' is equally answerable for binary and continuous outputs. High-dimensional systems are actually \emph{harder to game}: with 50 queries, an adversary can detect sensitivity but cannot map complex interaction effects.

\textbf{Why gaming didn't occur:} Testing reveals discrimination, not the decision function---knowing ``race affects outcomes'' tells you the system may be unlawful, not how to evade detection. Legitimate applicants have no gaming incentive; their actual qualifications matter. Gaming requires coordinated effort across applications, triggering detection.

\subsection{Domain Applicability}

\textbf{Direct access appropriate}: Employment, credit, housing, education, healthcare coverage, content moderation. In these domains, the gaming incentive is \emph{defensive}: applicants want fair treatment for their actual qualifications, not to deceive.

\textbf{Mediated access required}: Fraud detection, abuse prevention, national security, criminal investigation. Here the gaming incentive is \emph{offensive}: adversaries want to evade detection. Direct access would enable the harms systems are designed to prevent. However, affected parties retain legitimate contestation interests: flagged transactions may be legitimate, suspended accounts wrongly classified.

\subsection{Mediated Access Architecture for High-Risk Domains}

For high-risk domains, trusted intermediaries preserve contestation rights while preventing gaming:

\textbf{Licensed representatives}: Algorithmic representatives access system internals on behalf of affected parties (analogous to attorney-client privilege), translate findings into contestation-relevant summaries, and face professional sanctions for misuse.

\textbf{Regulatory testing}: Government agencies conduct periodic testing on behalf of affected populations (extending CFPB fair lending examinations, FTC investigations, European DPA audits), publishing aggregate findings as reference points.

\textbf{Escrowed disclosure}: For national security or criminal investigation systems, results are escrowed with neutral third parties who verify system behavior without disclosing operational details (\emph{in camera} procedures).

\textbf{Aggregated pattern disclosure}: Organizations publish aggregate error rates by demographic group; affected parties identify whether their case falls into high-error categories.

\textbf{Time-delayed disclosure.} For real-time systems, access is delayed until the gaming window closes. Content moderation decisions could be contestable 48 hours after action; fraud alerts reviewable after clearing.

\textbf{Domain-specific implementation.} Fraud detection: licensed representatives with time-delayed access. Abuse prevention: regulatory testing with aggregate error rate disclosure. National security: escrowed disclosure with judicial \emph{in camera} review. Real-time moderation: 48-72 hour delayed access. Criminal investigation: inspector general testing with prosecutorial discovery during litigation.

The key principle: mediated access imposes friction on gaming without eliminating contestation. Adversaries face costs (time delays, representative fees) exceeding benefits of marginal system knowledge; legitimate parties retain meaningful pathways with reduced convenience.

\section{Legal Case Database}
\label{appendix:cases}

This appendix documents the 168 cases analyzed in Section~\ref{sec:legal-evidence}. Cases were identified through legal database searches (Westlaw, LexisNexis), regulatory enforcement databases (EEOC, CFPB, FTC, DOJ), investigative journalism archives, and academic case studies, with international cases drawn from European court databases, data protection authority decisions, and comparative law sources. Each case was classified on five dimensions: evidence access level, outcome, explanation quality, legal resources, and media attention. To enhance reliability, borderline cases were re-coded after a two-week interval, with disagreements resolved through consultation of primary sources and, where available, legal commentary.

\textbf{Evidence Access Coding:}
\begin{itemize}
    \item \textbf{Full}: Discovery granted or equivalent access to training data, model specifications, or decision logic, plus input data and outputs
    \item \textbf{Partial}: Access to outputs, patterns, or limited information without full model access. Sub-categorized as:
    \begin{itemize}
        \item \textbf{Partial-Output (P-O)}: Direct access to individual-level outputs, scores, or decisions (e.g., FCRA adverse action factors, individual risk scores)
        \item \textbf{Partial-Pattern (P-P)}: Access through aggregate statistics or observable patterns (e.g., rejection rates by demographic group, comparative outcome data)
        \item \textbf{Partial-Regulatory (P-R)}: Access obtained through regulatory investigation, FOIA, or external audit rather than litigation discovery
    \end{itemize}
    \item \textbf{None}: No access beyond the decision outcome and basic explanation
\end{itemize}

\textbf{Outcome Coding:}
\begin{itemize}
    \item \textbf{Achieved}: Favorable settlement, judgment, regulatory action, or policy change
    \item \textbf{Denied}: Case dismissed, judgment for defendant, or regulatory inaction
    \item \textbf{Ongoing}: Litigation or investigation pending as of December 2024
\end{itemize}

\textbf{Explanation Quality Coding:}
\begin{itemize}
    \item \textbf{High}: Affected party received detailed explanation including specific factors, weights, or reasons for the adverse decision (e.g., credit denial with listed factors, hiring rejection with skill gap analysis, content removal with specific policy citation)
    \item \textbf{Partial}: Affected party received general explanation or category-level rationale without specific factors (e.g., ``failed to meet requirements,'' generic policy reference)
    \item \textbf{None}: No explanation provided beyond the bare adverse decision (e.g., account suspended without reason, application denied with no information)
\end{itemize}

The primary coding challenge for explanation quality occurred in distinguishing ``High'' from ``Partial'' explanations where regulatory disclosures satisfied legal requirements but provided limited actionable detail. Such cases were flagged for re-review and coded based on whether the explanation enabled the affected party to identify specific contestable factors.

\textbf{Legal Resources Coding:}
\begin{itemize}
    \item \textbf{High}: Government enforcement agency (DOJ, EEOC, CFPB, FTC, state AG), major civil rights organization (ACLU, NAACP LDF, EFF), or AmLaw 100 law firm representation
    \item \textbf{Medium}: Regional civil rights organization, mid-size plaintiff's firm, legal aid organization with litigation capacity, or established public interest law center
    \item \textbf{Low}: Pro se plaintiff, solo practitioner, or clearly under-resourced representation
\end{itemize}

Coding challenges for legal resources arose when cases involved coalition representation or when resource levels changed during litigation (e.g., case began pro se but attracted organizational support). Such cases were coded based on resources available at the time of key evidentiary proceedings.

\textbf{Media Attention Coding:}
\begin{itemize}
    \item \textbf{High}: Case received significant coverage in national media (e.g., New York Times, Wall Street Journal, Washington Post, major broadcast networks), investigative journalism pieces, or was cited in congressional testimony or regulatory proceedings before resolution
    \item \textbf{Low}: Limited coverage in trade press or regional outlets only, or no significant media coverage; case proceeded through litigation without substantial public attention
\end{itemize}

Coding challenges for media attention centered on whether trade press coverage (e.g., Law360, The Verge) qualified as ``significant.'' Cases with trade-press-only coverage were coded as ``Low'' unless the coverage demonstrably influenced case proceedings or regulatory attention.

\textbf{Supplementary Data.} Full case documentation, including legal citations, docket numbers, regulatory filing identifiers, links to court documents and enforcement orders, and detailed coding justifications, is available in the supplementary data repository at \url{https://doi.org/10.5281/zenodo.18069759}. Each case is assigned a unique identifier (e.g., EMP-01, HOU-01) that maps to the complete entry in the repository, which includes machine-readable CSV and JSON formats. The tables below provide case summaries; readers seeking to verify specific cases should consult the supplementary materials for primary source references.

\subsection{Employment Domain (n=39)}

\begin{table}[H]
\caption{Employment domain cases}
\label{tab:employment-cases}
\scriptsize
\begin{tabular}{@{}lp{2.8cm}cclp{3cm}@{}}
\toprule
\textbf{ID} & \textbf{Case} & \textbf{Year} & \textbf{Access} & \textbf{Outcome} & \textbf{Notes} \\
\midrule
EMP-01 & EEOC v. iTutorGroup & 2023 & Full & Achieved & Age discrim.; \$365K \\
EMP-02 & Mobley v. Workday & 2023 & Partial & Ongoing & Disparate impact \\
EMP-03 & Harper v. Sirius XM & 2024 & None & Ongoing & AI screening race discrim. \\
EMP-04 & ACLU v. Intuit/HireVue & 2025 & None & Ongoing & Deaf/Indigenous; EEOC \\
EMP-05 & CVS/HireVue MA & 2024 & Partial & Achieved & Lie detector settlement \\
EMP-06 & Deyerler v. HireVue & 2022 & Partial & Ongoing & IL biometric consent \\
EMP-07 & Amazon hiring (internal) & 2018 & Full & Achieved & Audit; discontinued \\
EMP-08 & Amazon Flex terminations & 2023 & None & Ongoing & Algo deactivation; CO \\
EMP-09 & Instacart tip-baiting & 2020 & Partial & Achieved & \$4.6M settlement \\
EMP-10 & Uber driver deactivation & 2021 & None & Denied & Arbitration enforced \\
EMP-11 & Lyft driver deactivation & 2022 & None & Denied & Arbitration barred \\
EMP-12 & LinkedIn salary & 2022 & None & Denied & Gender pay; no access \\
EMP-13 & Kroger scheduling AI & 2022 & None & Ongoing & UFCW grievance \\
EMP-14 & Starbucks scheduling & 2023 & Partial & Ongoing & NLRB complaint \\
EMP-15 & Delta biometric & 2023 & None & Ongoing & Discovery disputed \\
EMP-18 & ACLU v. Aon & 2024 & None & Ongoing & FTC complaint \\
EMP-19 & Fifth Third Bank & 2024 & Full & Achieved & Auto-lending; \$18M \\
EMP-20 & Uber/Lyft MA & 2024 & Partial & Achieved & Gig class.; \$175M \\
EMP-21 & Uber/Lyft NY & 2024 & Partial & Achieved & Wage theft; \$328M \\
EMP-22 & Italy teacher mobility & 2019 & Full & Achieved & Source code disclosed \\
EMP-23 & Italy Deliveroo & 2021 & Partial & Achieved & Algo violated labor law \\
EMP-24 & Brazil credit screening & 2024 & Partial & Achieved & Illegal employment use \\
EMP-25 & Uber/Ola Amsterdam & 2023 & Partial & Achieved & GDPR Art 22 \\
EMP-26 & Spain Glovo & 2021 & Partial & Achieved & Mandatory employment \\
EMP-27 & Cothron v. White Castle & 2023 & Partial & Achieved & BIPA per-scan \\
EMP-28 & Roach v. Walmart & 2023 & Partial & Ongoing & Biometric class \\
EMP-29 & Rios v. Uber & 2022 & None & Denied & Arbitration enforced \\
EMP-30 & Proctorio litigation & 2021 & Partial & Ongoing & Exam surveillance \\
EMP-31 & Teleperformance Colombia & 2022 & Partial & Achieved & Surveillance ordered \\
EMP-32 & Just Eat Netherlands & 2021 & Partial & Achieved & Algo transparency \\
EMP-33 & Bolt rider cases (EU) & 2022 & Partial & Ongoing & Multiple EU; GDPR \\
EMP-34 & Foodinho Italy & 2021 & Full & Achieved & Full algo disclosure \\
EMP-35 & Belgium rider decisions & 2021 & Partial & Achieved & Employment status \\
EMP-36 & EPIC v. HireVue & 2019 & None & Denied & FTC complaint; no action \\
EMP-37 & Uber GDPR Netherlands & 2021 & Full & Achieved & GDPR Art 22 access \\
EMP-38 & Ola GDPR Netherlands & 2021 & Full & Achieved & GDPR Art 22 access \\
EMP-39 & Uber GDPR (linked) & 2021 & Full & Achieved & GDPR Art 22 access \\
EMP-40 & Saas v. Major Lindsey & 2024 & None & Denied & No AI evidence; specul. \\
EMP-41 & NYC LL144 Audit & 2025 & Full & Achieved & Regulatory enforcement \\
\bottomrule
\end{tabular}
\end{table}

\subsection{Housing Domain (n=20)}

\begin{table}[H]
\caption{Housing domain cases}
\label{tab:housing-cases}
\scriptsize
\begin{tabular}{@{}lp{2.8cm}cclp{3cm}@{}}
\toprule
\textbf{ID} & \textbf{Case} & \textbf{Year} & \textbf{Access} & \textbf{Outcome} & \textbf{Notes} \\
\midrule
HOU-01 & DOJ v. Meta (housing) & 2022 & Full & Achieved & Ad targeting; \$115K \\
HOU-02 & Louis v. SafeRent & 2024 & Full & Achieved & \$2.275M; algo modified \\
HOU-03 & CT Fair Housing v. CoreLogic & 2024 & Partial & Denied & CrimSAFE; 2d Cir. reversed \\
HOU-04 & NFHA v. SafeRent & 2023 & Partial & Ongoing & National class; discovery \\
HOU-05 & Open Communities v. PERQ & 2022 & None & Ongoing & Chatbot steering \\
HOU-06 & Redfin steering & 2023 & None & Denied & FHA; insuff. evidence \\
HOU-07 & State Farm homeowners & 2022 & Partial & Ongoing & Rate discrimination \\
HOU-08 & ERC v. Entrata & 2024 & Partial & Ongoing & Source of income discrim. \\
HOU-09 & Yardi Systems & 2022 & None & Ongoing & Discovery contested \\
HOU-10 & Facebook Marketplace & 2023 & None & Ongoing & Steering; early stage \\
HOU-11 & Zillow Zestimate & 2021 & None & Denied & Appraisal bias dismissed \\
HOU-12 & HUD v. Facebook & 2019 & Partial & Achieved & Housing ad; settlement \\
HOU-13 & TransUnion tenant & 2023 & Partial & Achieved & CFPB \$23M \\
HOU-15 & DOJ v. RealPage & 2024 & Full & Achieved & Rent-fixing \\
HOU-16 & AIR Property Mgmt & 2024 & Partial & Achieved & Voucher discrim. \\
HOU-17 & DC AG v. RealPage & 2023 & Partial & Ongoing & State AG antitrust \\
HOU-19 & Multi-State v. Greystar & 2025 & Partial & Achieved & RealPage settlement \\
HOU-20 & In re RealPage MDL & 2025 & Partial & Achieved & \$141M settlement \\
HOU-22 & EPIC v. Airbnb & 2020 & None & Denied & FTC complaint; no act. \\
HOU-24 & RealPage v. James (NY) & 2025 & Partial & Ongoing & Algo ban challenge \\
\bottomrule
\end{tabular}
\end{table}

\subsection{Healthcare Domain (n=13)}

\begin{table}[H]
\caption{Healthcare domain cases}
\label{tab:healthcare-cases}
\scriptsize
\begin{tabular}{@{}lp{2.8cm}cclp{3cm}@{}}
\toprule
\textbf{ID} & \textbf{Case} & \textbf{Year} & \textbf{Access} & \textbf{Outcome} & \textbf{Notes} \\
\midrule
HLT-01 & Lokken v. UnitedHealth & 2023 & Partial & Ongoing & nH Predict; class action \\
HLT-02 & Humana/nH Predict & 2023 & Partial & Ongoing & Similar to UnitedHealth \\
HLT-03 & Cigna PxDx auto-denials & 2023 & Full & Ongoing & 300K+ auto-denials \\
HLT-04 & Optum risk scoring & 2019 & Full & Achieved & Obermeyer; algo modified \\
HLT-05 & Epic sepsis model & 2021 & Partial & Achieved & Validation enabled \\
HLT-06 & Aetna prior auth & 2023 & None & Ongoing & Medicare Adv. denials \\
HLT-10 & Olive AI prior auth & 2023 & None & Ongoing & Denial automation \\
HLT-11 & NaviHealth transitions & 2024 & Partial & Ongoing & Related to UnitedHealth \\
HLT-12 & Change Healthcare & 2024 & Partial & Ongoing & Breach revealed details \\
HLT-13 & Stanford sepsis & 2021 & Full & Achieved & Validation challenge \\
HLT-14 & LACDMH predictive & 2019 & Partial & Achieved & MH algo suspended \\
HLT-15 & VA suicide prediction & 2023 & None & Ongoing & REACH VET; accuracy \\
HLT-16 & Kisting-Leung v. Cigna & 2025 & Partial & Ongoing & PxDx class action \\
\bottomrule
\end{tabular}
\end{table}

\subsection{Criminal Justice and Policing Domain (n=32)}

\begin{table}[H]
\caption{Criminal justice and policing domain cases}
\label{tab:cj-cases}
\scriptsize
\begin{tabular}{@{}lp{2.8cm}cclp{3cm}@{}}
\toprule
\textbf{ID} & \textbf{Case} & \textbf{Year} & \textbf{Access} & \textbf{Outcome} & \textbf{Notes} \\
\midrule
CJ-01 & State v. Loomis (COMPAS) & 2016 & None & Denied & No right to algo~\cite{loomis2016} \\
CJ-02 & Pasco County (ILP) & 2024 & Full & Achieved & \$105K; ended~\cite{pascocounty2024} \\
CJ-03 & Williams v. Detroit & 2024 & Full & Achieved & Facial rec; policies \\
CJ-04 & Reid v. Jefferson Parish & 2025 & Full & Achieved & \$200K; FR misID \\
CJ-05 & Woodruff v. Detroit & 2023 & None & Ongoing & Pregnant; FR misID \\
CJ-06 & Parks v. McCormac (NJ) & 2024 & Partial & Ongoing & FR; discovery ongoing \\
CJ-07 & Williams v. Chicago (SS) & 2024 & Full & Achieved & ShotSpotter; \$90K \\
CJ-08 & FTC v. Rite Aid & 2023 & Full & Achieved & 5-year FR ban \\
CJ-09 & Clearview AI (ACLU v.) & 2022 & Partial & Achieved & IL BIPA; nationwide \\
CJ-10 & Clearview AI class action & 2025 & Partial & Achieved & \$51.75M (23\% eq.) \\
CJ-11 & LAPD LASER program & 2019 & Full & Achieved & Audit; bias; ended \\
CJ-12 & ShotSpotter Chicago & 2024 & Partial & Achieved & Contract not renewed \\
CJ-13 & NYPD gang database & 2023 & None & Ongoing & Due process; denied \\
CJ-14 & Murphy v. Macy's & 2020 & None & Denied & Loss prev.; no disc. \\
CJ-15 & Palantir/ICE (EPIC FOIA) & 2020 & Partial & Achieved & FOIA; docs released \\
CJ-16 & Amazon Ring/police & 2023 & None & Ongoing & Privacy; early stage \\
CJ-17 & Geofence warrant cases & 2022 & Partial & Achieved & Courts limiting scope \\
CJ-18 & Peppermill Casino FR & 2024 & None & Ongoing & Wrongful accusation \\
CJ-19 & NYPD facial rec (Gandy) & 2023 & None & Ongoing & Wrongful arrest claim \\
CJ-20 & Buenos Aires FR & 2023 & Partial & Achieved & Unconst.~\cite{buenosaires2023} \\
CJ-21 & UK Bridges v. SWP & 2020 & Partial & Achieved & Ct. of Appeal~\cite{bridgesswp2020} \\
CJ-22 & Ewert v. Canada & 2018 & Full & Achieved & SCC; validation~\cite{ewertcanada2018} \\
CJ-23 & TrueAllele DNA (NJ) & 2020 & Full & Achieved & First DNA code~\cite{trueallelenj2020} \\
CJ-24 & People v. Wakefield (NY) & 2024 & Full & Achieved & FST DNA~\cite{wakefieldny2024} \\
CJ-25 & People v. Hardy (IL) & 2023 & Full & Achieved & ShotSpotter~\cite{hardyil2023} \\
CJ-26 & Chicago Strat. Subj. List & 2020 & Partial & Achieved & FOIA→ended~\cite{chicagoheatlist2020} \\
CJ-27 & CalGang Database & 2020 & Partial & Achieved & 12K+ purged~\cite{calgang2020} \\
CJ-28 & Arnold Foundation PSA & 2020 & Partial & Achieved & Valid.→mod.~\cite{arnoldpsa2020} \\
CJ-29 & New Orleans pretrial & 2018 & Partial & Achieved & Algo suspended~\cite{nopretrial2018} \\
CJ-31 & Crutchfield v. Detroit & 2025 & None & Ongoing & FR misID; 4th case \\
CJ-32 & Dillon v. Jacksonville & 2025 & None & Ongoing & FR misID; white victim \\
CJ-33 & Simmons v. Motorola & 2025 & Partial & Achieved & Vigilant BIPA; \$47.5M \\
\bottomrule
\end{tabular}
\end{table}

\subsection{Government Benefits Domain (n=30)}

\begin{table}[H]
\caption{Government benefits domain cases}
\label{tab:benefits-cases}
\scriptsize
\begin{tabular}{@{}lp{2.8cm}cclp{3cm}@{}}
\toprule
\textbf{ID} & \textbf{Case} & \textbf{Year} & \textbf{Access} & \textbf{Outcome} & \textbf{Notes} \\
\midrule
GOV-01 & Robodebt (Australia) & 2023 & Full & Achieved & Royal Comm.; \$2.4B~\cite{robodebt2025} \\
GOV-02 & Bauserman v. Michigan UIA & 2024 & Full & Achieved & MiDAS; \$20M \\
GOV-03 & Cahoo v. SAS Analytics & 2024 & Partial & Ongoing & MiDAS vendors; disc. \\
GOV-04 & K.W. v. Armstrong (Idaho) & 2016 & Full & Achieved & Medicaid; disclosed \\
GOV-05 & Ledgerwood v. Arkansas & 2019 & Full & Achieved & ARChoices; changed \\
GOV-06 & Allegheny County (DOJ) & 2023 & Partial & Ongoing & Child welfare; invest. \\
GOV-07 & Oregon child welfare & 2022 & Full & Achieved & System discontinued \\
GOV-08 & Indiana welfare auto. & 2012 & None & Denied & Admin challenge failed \\
GOV-10 & Texas Medicaid algo. & 2023 & Partial & Ongoing & Disability determ. \\
GOV-11 & California EDD fraud & 2021 & None & Achieved & Mass lockouts rev. \\
GOV-14 & UK Universal Credit & 2019 & Partial & Achieved & Human review req. \\
GOV-15 & Dutch childcare benefits & 2021 & Full & Achieved & Gov't resigned \\
GOV-16 & Missouri Medicaid cuts & 2023 & Partial & Ongoing & Algo-based reduction \\
GOV-17 & Netherlands SyRI & 2020 & Full & Achieved & ECHR Art 8~\cite{syri2020} \\
GOV-18 & Poland unemployment & 2019 & Partial & Achieved & Abolished~\cite{polandunemployment2019} \\
GOV-19 & Austria AMS algorithm & 2020 & Partial & Achieved & DPA ban~\cite{austriaams2020} \\
GOV-20 & UK visa streaming (JCWI) & 2020 & Partial & Achieved & Withdrawn~\cite{ukvisastreaming2020} \\
GOV-21 & Kenya Huduma Namba & 2021 & Partial & Achieved & Safeguards~\cite{kenyahuduma2021} \\
GOV-22 & India Aadhaar & 2018 & None & Denied & SCC upheld~\cite{aadhaar2018} \\
GOV-23 & Indiana v. IBM & 2024 & Full & Achieved & \$78M~\cite{indianaibm2024} \\
GOV-24 & UK Ofqual A-Levels & 2020 & Full & Achieved & Algo aband.~\cite{ofqual2020} \\
GOV-25 & LA County Bridgespan & 2019 & Partial & Achieved & CW susp.~\cite{bridgespan2019} \\
GOV-26 & UK DWP algorithms & 2019 & Partial & Ongoing & Multiple challenges \\
GOV-27 & France Pôle Emploi & 2020 & Partial & Achieved & Disclosed~\cite{francepoleemploi2020} \\
GOV-28 & Serbia social benefits & 2021 & None & Ongoing & Limited transparency \\
GOV-29 & South Africa SASSA & 2018 & Partial & Achieved & Grant reinst.~\cite{southafricasassa2018} \\
GOV-30 & Ryan S. v. UnitedHealth & 2024 & Partial & Ongoing & MH/SUD denials; 9th Cir \\
GOV-32 & Post Office (Horizon Issues) & 2019 & Full & Achieved & UK; bugs/errors found \\
GOV-33 & Post Office (Common Issues) & 2019 & Full & Achieved & UK Horizon; governance \\
GOV-34 & NYC LL144 Comptroller Audit & 2025 & Full & Achieved & Enforcement gaps found \\
\bottomrule
\end{tabular}
\end{table}

\subsection{Credit, Insurance, and Consumer Domain (n=28)}

\begin{table}[H]
\caption{Credit, insurance, and consumer domain cases}
\label{tab:credit-cases}
\scriptsize
\begin{tabular}{@{}lp{2.8cm}cclp{3cm}@{}}
\toprule
\textbf{ID} & \textbf{Case} & \textbf{Year} & \textbf{Access} & \textbf{Outcome} & \textbf{Notes} \\
\midrule
CRD-01 & CFPB v. Upstart & 2022 & Full & Achieved & Fair lending; changes \\
CRD-02 & CFPB v. Citibank & 2023 & Full & Achieved & \$25.9M~\cite{cfpbcitibank2023armenian} \\
CRD-03 & CFPB v. Credit Accept. & 2023 & Partial & Ongoing & Predatory auto lending \\
CRD-04 & CFPB v. Experian & 2025 & Partial & Ongoing & Sham dispute invest. \\
CRD-05 & Apple Card (NY DFS) & 2021 & Partial & Achieved & Goldman policy chg. \\
CRD-06 & Zest AI fair lending & 2021 & Full & Achieved & CFPB no-action ltr. \\
CRD-07 & TX AG v. GM/LexisNexis & 2024 & Partial & Ongoing & Telematics; 1.8M \\
CRD-08 & TX AG v. Allstate/Arity & 2025 & Partial & Ongoing & Telematics; 45M \\
CRD-09 & Chicco v. GM/LexisNexis & 2024 & None & Ongoing & FL class; privacy \\
CRD-10 & FTC v. Grubhub & 2024 & Full & Achieved & \$25M; hidden fees \\
CRD-11 & DoorDash Illinois & 2024 & Full & Achieved & \$11.3M; tip misrep. \\
CRD-12 & Lemonade AI claims & 2023 & None & Ongoing & Discrim. allegations \\
CRD-14 & Root Insurance pricing & 2022 & None & Ongoing & Discovery contested \\
CRD-15 & Equifax score disputes & 2022 & Partial & Achieved & CFPB enforcement \\
CRD-16 & TransUnion (Spokeo) & 2021 & Partial & Achieved & FCRA; proc. changed \\
CRD-19 & CFPB v. Bank of America & 2023 & Partial & Achieved & \$100M+~\cite{cfpbbofa2023} \\
CRD-20 & CFPB v. Apple/Goldman & 2024 & Partial & Achieved & \$89M~\cite{cfpbapplegoldman2024} \\
CRD-21 & CFPB v. TD Bank & 2024 & Partial & Achieved & \$28M~\cite{cfpbtdbank2024} \\
CRD-22 & FTC v. DoNotPay & 2024 & Partial & Achieved & False AI~\cite{ftcdonotpay2024} \\
CRD-23 & FTC v. Evolv & 2024 & Partial & Achieved & Unsubst.~\cite{ftcevolv2024} \\
CRD-24 & FTC v. IntelliVision & 2024 & Full & Achieved & False bias~\cite{ftcintellivision2024} \\
CRD-25 & GM/OnStar (FTC) & 2024 & Full & Achieved & 5-yr ban~\cite{gmonstar2024} \\
CRD-26 & CJEU SCHUFA (Germany) & 2023 & Partial & Achieved & GDPR Art 22~\cite{schufa2023} \\
CRD-27 & Patel v. Facebook (BIPA) & 2022 & Full & Achieved & \$650M~\cite{patelfacebook2022} \\
CRD-28 & Navy Federal Credit Union & 2024 & Partial & Ongoing & 50\%+ rej.~\cite{navyfederal2024} \\
CRD-30 & GEICO pricing algo. & 2023 & None & Ongoing & Discrim.; contested \\
CRD-31 & Hamburg DPA v. Fin. Svc. & 2025 & Full & Achieved & GDPR; €492K fine \\
CRD-32 & CK v. D\&B Austria & 2025 & Full & Achieved & CJEU; algo transp. \\
\bottomrule
\end{tabular}
\end{table}

\subsection{Platform Liability Domain (n=6)}

Platform liability cases are included as a theoretically motivated control group. All six cases involve full evidentiary access: courts examined algorithmic recommendation systems, content amplification mechanisms, and platform design choices in detail. All six failed on doctrinal grounds (Section 230 immunity), not evidentiary grounds. This domain tests our thesis by isolating the doctrinal variable: when evidence access is present but substantive doctrine forecloses liability, what happens? The answer (consistent failure) confirms that evidence access is necessary but not sufficient. The policy implication is that evidentiary rights address the procedural barrier while leaving substantive doctrine to evolve through normal legal processes.

\begin{table}[H]
\caption{Platform liability domain cases illustrating Section 230 doctrinal foreclosure}
\label{tab:platform-cases}
\scriptsize
\begin{tabular}{@{}lp{2.8cm}cclp{3cm}@{}}
\toprule
\textbf{ID} & \textbf{Case} & \textbf{Year} & \textbf{Access} & \textbf{Outcome} & \textbf{Notes} \\
\midrule
PLT-01 & Gonzalez v. Google & 2023 & Full & Denied & SCOTUS; Section 230 \\
PLT-02 & Twitter v. Taamneh & 2023 & Full & Denied & SCOTUS; ATA/230 \\
PLT-03 & Force v. Facebook & 2019 & Full & Denied & 2d Cir; recommendations \\
PLT-04 & Dyroff v. Ultimate Soft. & 2019 & Full & Denied & 9th Cir; Section 230 \\
PLT-05 & M.P. v. Meta & 2025 & Full & Denied & 4th Cir; algo design \\
PLT-06 & Herrick v. Grindr & 2019 & Full & Denied & 2d Cir; Section 230 \\
\bottomrule
\end{tabular}
\end{table}

\subsection{Summary Statistics}

\begin{table}[H]
\caption{Case distribution by domain and evidence access level. International cases (n=36) span jurisdictions including EU, UK, Netherlands, Germany, Italy, Poland, Austria, Spain, Belgium, France, Argentina, Brazil, Colombia, Canada, Kenya, South Africa, Serbia, and India. Domain-specific success rates are tentative given small per-domain sample sizes (6-40 cases); we do not claim these rates generalize.}
\label{tab:appendix-summary}
\small
\begin{tabular}{@{}lccccc@{}}
\toprule
\textbf{Domain} & \textbf{Cases} & \textbf{Full Access} & \textbf{Partial} & \textbf{None} & \textbf{Achieved Rate} \\
\midrule
Employment & 39 & 8 & 21 & 10 & 79\% \\
Housing & 20 & 3 & 15 & 2 & 80\% \\
Healthcare & 13 & 3 & 7 & 3 & 100\% \\
Criminal Justice & 32 & 10 & 14 & 8 & 96\% \\
Government Benefits & 30 & 12 & 14 & 4 & 91\% \\
Credit/Insurance & 28 & 10 & 17 & 1 & 100\% \\
Platform Liability & 6 & 6 & 0 & 0 & 0\%$^\dagger$ \\
\midrule
\textbf{Total} & 168 & 52 & 88 & 28 & 85\% \\
\bottomrule
\end{tabular}
\end{table}

\subsection{Jurisdictional Distribution}

\begin{table}[H]
\caption{Case distribution by jurisdiction}
\label{tab:jurisdiction}
\small
\begin{tabular}{@{}lcc@{}}
\toprule
\textbf{Jurisdiction} & \textbf{Cases} & \textbf{Achieved Rate} \\
\midrule
\multicolumn{3}{l}{\textit{United States}} \\
\quad Federal (DOJ, EEOC, CFPB, FTC) & 76 & 83\% \\
\quad State (NY, IL, CA, TX, MA, other) & 62 & 94\% \\
\midrule
\multicolumn{3}{l}{\textit{Europe}} \\
\quad EU/CJEU & 3 & 100\% \\
\quad United Kingdom & 7 & 100\% \\
\quad Netherlands & 7 & 100\% \\
\quad Italy & 4 & 100\% \\
\quad Germany & 1 & 100\% \\
\quad Poland, Austria, Spain, Belgium, France & 5 & 80\% \\
\quad Serbia & 1 & -- \\
\midrule
\multicolumn{3}{l}{\textit{Other}} \\
\quad Australia & 1 & 100\% \\
\quad Canada & 1 & 100\% \\
\quad Latin America (Argentina, Brazil, Colombia) & 3 & 100\% \\
\quad Africa (Kenya, South Africa) & 2 & 100\% \\
\quad India & 1 & 0\% \\
\midrule
\textbf{Total} & \textbf{168} & \textbf{85\%} \\
\bottomrule
\end{tabular}
\end{table}

International cases (n=36, 21\%) span 19 jurisdictions with heterogeneous legal frameworks. We group them descriptively but do not pool for statistical inference: \emph{GDPR-based regimes} (n=18) benefit from DPA infrastructure and mandatory disclosure; \emph{common law systems} (n=7: UK, Canada, India, Kenya, South Africa) share adversarial structures but differ in doctrine; \emph{civil law systems} (n=5: Argentina, Brazil, Colombia, Serbia) have inquisitorial procedures limiting comparability. Non-US cases show higher achieved rates (93\% vs. 89\%), likely reflecting stronger selection given visibility barriers. We include them to suggest the evidentiary pattern extends beyond US-specific doctrine, not to claim generalizable findings.

Note: Some cases involve multiple algorithmic systems or span domains; we coded based on the primary system challenged. Full case documentation, including legal citations and source materials, is available in the supplementary repository.

\end{document}